\documentclass{aa}
\usepackage{epsfig}
\usepackage{graphicx}
\usepackage{mathtools}
\usepackage{txfonts}
\usepackage{lscape}
\usepackage{hyperref}

\bibpunct{(}{)}{;}{a}{}{,}

\begin{document}

\title{
Inferring the three-dimensional distribution of dust in the Galaxy with a non-parametric method
}
\subtitle{Preparing for Gaia}

\author{S. Rezaei Kh. \and C.A.L. Bailer-Jones \and R.J. Hanson \and M. Fouesneau\\
}
\institute{Max Plank Institute for Astronomy (MPIA),  K\"onigstuhl 17, 69117 Heidelberg, Germany}

\def\teff{$T_{\rm eff}$}

\def\expec{{\mathrm E}}
\def\cov{{\mathrm Cov}}

\def\trans{^\mathsf{T}}
\def\inv{^{-1}}
\def\mby{\!\times\!}
\def\mplus{\!+\!}

\def\geonj{g_{n,j}}
\def\gvec{{\mathbf g}}
\def\geomat{{\mathrm G}}

\def\likecovN{{\mathrm V}_N}

\def\extn{a_n}
\def\extsdn{\sigma_n}
\def\extvecN{{\mathbf a}_N}
\def\extcovN{\Sigma_N}
\def\estextn{f_n}
\def\estextvecN{{\mathbf f}\!_N}

\def\rhonj{\rho_{n,j}}
\def\rhovecJ{{\boldsymbol \rho}_J}
\def\rhovecJp{{\boldsymbol \rho}_{J+1}}
\def\rhoJp{\rho_{J+1}}

\def\gpcov{{\mathrm C}}
\def\gpcovJ{{\mathrm C}_J}
\def\gpcovJp{{\mathrm C}_{J+1}}
\def\gpcovel{c}
\def\kvecJ{{\mathbf k}_J}
\def\kJp{k}

\def\mmatJ{{\mathrm M}_J}
\def\mvecJ{{\mathbf m}_J}

\def\bvecJ{{\mathbf b}_J}
\def\hvecJ{{\mathbf h}_J}
\def\rmatJ{{\mathrm R}_J}

\def\rvec{{\mathbf r}}

\def\ofo{{\mathcal O}}

\abstract{We present a non-parametric model for inferring the three-dimensional (3D) distribution of dust density in the Milky Way. Our approach uses the extinction measured towards stars at different locations in the Galaxy at approximately known distances. Each extinction measurement is proportional to the integrated dust density along its line of sight (l.o.s). Making simple assumptions about the spatial correlation of the dust density, we can infer the most probable 3D distribution of dust across the entire observed region, including along sight lines which were not observed.  This is possible because our model employs a Gaussian process to connect all l.o.s. We demonstrate the capability of our model to capture detailed dust density variations using mock data and simulated data from the Gaia Universe Model Snapshot.  We then apply our method to a sample of giant stars observed by APOGEE and Kepler to construct a 3D dust map over a small region of the Galaxy. Owing to our smoothness constraint and its isotropy, we provide one of the first maps which does not show the ``fingers of God'' effect.  }


\maketitle

\section{introduction}

Interstellar dust is an integral part of a galaxy. In the cycle of matter, gas and dust are ejected from evolving stars, and eventually this material will form new stars.  Dust also attenuates light through reddening and extinction, thereby complicating our interpretation of stellar photometry and spectroscopy, and it imposes complex selection functions on surveys.  Any survey with or of stars must account for the effects of interstellar dust.

Numerous studies have been undertaken over the years to improve our knowledge of this component of the interstellar medium (ISM). Early extinction maps were emission-based and in two dimensions (2D). A prominent piece of work was that of \citet{1998ApJ...500..525S} who used far-infrared dust emission measured by the IRAS and COBE satellites to build an all-sky map of the dust column density, assuming a standard reddening law.
\citet{2006A&A...454..781L} used 2MASS photometric data and the colour excess technique to study molecular clouds and to map the dust column density and extinction in the Pipe nebula. \citet{2006A&A...453..635M} used a Galaxy model to find the intrinsic colours of stars, then, using the measured near infrared colour excess, estimated the extinction and distances towards stars. They applied their model to more than 64\,000 l.o.s to find the 3D distribution of extinction in the inner Galaxy. \citet{2010ApJ...725.1175S} used the Sloan Digital Sky Survey (SDSS) and the blue tip of the stellar locus to measure the colour shift of the main sequence turnoff and thereby calculate the reddening of stars.

\citet{2012MNRAS.427.2119S} introduced a hierarchical Bayesian model to simultaneously estimate a distance--extinction relation and the properties of individual stars from multi-band photometry. This improved the precision and accuracy of the maps compared to previous ones. This method was used by \citet{2014MNRAS.443.2907S} to build a 3D extinction map of the northern Galactic plane from
IPHAS photometry. A 3D map of interstellar dust reddening for three-quarters of the sky was presented by \citet{2015ApJ...810...25G} using Pan-STARRS1 and 2MASS photometry, based on the method of \citet{2014ApJ...783..114G} (which uses a Bayesian approach similar to that of \citealt{2012MNRAS.427.2119S}). This method was also used by \citet{2014ApJ...789...15S} to make a map of dust reddening out to 4.5 kpc from Pan-STARRS1 stellar photometry which covers the entire sky north of declination -30$^{\circ}$. Their method was especially designed for modelling extinction in the Galactic plane.

A similar approach was taken by \citet{2014MNRAS.438.2938H}, who used SDSS and UKIDSS multi-band photometry to map extinction in 3D. They used a Bayesian model to take into account the degeneracy between extinction and stellar effective temperature. The method was previously introduced by \citet{2011MNRAS.411..435B} to estimate effective temperatures and reddenings towards 40\,000 FGK stars from 2MASS and Hipparcos photometry and Hipparcos parallaxes.  \citet{2014A&A...561A..91L} applied an inversion method to measurements of stellar colour excess made at optical wavelengths. Together with parallaxes or photometric distances they constructed a map of the ISM within 2.5 kpc of the Sun.  \citet{2014MNRAS.445..256S} presented a mapping method in which extinction is modelled as a Gaussian random field with a covariance function which has a Kolmogorov-like power spectrum. This is motivated by the idea that turbulence is responsible for the spatial structure of the ISM.

Many of the aforementioned methods consider each l.o.s independently: they do not propagate information between neighbouring l.o.s, even though the dust causing the extinction is likely to be correlated. This produces discontinuities in many published extinction maps, similar to an artefact known as the ``fingers of god''. In this paper, we present a method which uses extinctions and distances towards multiple stars in a collective manner simultaneously, by taking into account the correlation between neighbouring l.o.s; unlike earlier work by \citet{2014MNRAS.445..256S}, the covariance of our Gaussian process prior is in the dust density space; not in the extinction. This enables us to build a 3D map of the dust density which is free of the fingers-of-god effect, as a result of the isotropy of our smooth prior.  We do this by modelling the dust density, which is a local property of the ISM, rather than the extinction, which is the integral over all the dust along the l.o.s. We use a non-parametric model which allows us to avoid adopting an explicit -- and inevitably overly simple -- functional form for the variation of dust density in the Galaxy. The true variation of dust is far too complex to be captured by a parametric model, which we can both define in advance and fit well enough with available data.  We instead use a Gaussian process, which constrains the variation of the dust densities without choosing a particular functional form for its spatial variation. The Gaussian process instead defines the form of the covariance function between all points in space. This approach permits a wide range of functional variations in 3D space.

To build the dust map, we need the extinctions toward and distances of a large number of stars.
Such data are being obtained by Gaia, which will soon provide astrometry and spectrophotometry for more than $10^9$ stars brighter than a G-band magnitude of about 20.
The expected end-of-mission parallax standard error is around 25 $\mu$as at $G=15$ mag \citep{2014EAS....67...23D}, from which we can estimate distances \citep[e.g.][]{2015PASP..127..994B}. Gaia is also equipped with two low resolution spectrophotometers which will provide the spectral energy distribution of all observed sources.
The Gaia Data Processing and Analysis Consortium (DPAC)
will use these to estimate the astrophysical parameters of individual stars, including the l.o.s reddening/extinction \citep{2013A&A...559A..74B}. 
The end-of-mission Gaia $A_{V}$ precision is expected to be around 0.03 to 0.05 mag for stars with $G \leq 15$, increasing to 0.2 mag for sources down to $G=20$ (Andrae et al.\ 2016, in preparation).
Individual stellar extinctions and distances are therefore the inputs for our method.

This paper is organized as follows. We introduce our method in section \ref{method}, whilst covering some technical details in the appendix. In section \ref{simulations} we demonstrate the model and its ability to recover the true dust values, using both toy simulations and a a simulated Gaia catalogue. We apply the model to real data over a small region of the sky (the Kepler field) in section \ref{real_data}. We summarize in section \ref{discussion} and discuss some aspects of our method including its current strengths and weaknesses.

\section{Method}\label{method}

We wish to determine the 3D spatial distribution of dust given measurements of the l.o.s extinction caused by this dust toward a number of stars. Specifically, given these extinction measurements, we would like to find the probability distribution over the dust density at {\em any} point in space, and not necessarily a point along the l.o.s to one of these stars. 

\subsection{Problem setup}

\begin{figure} 
\begin{center}
\includegraphics[width=0.37\textwidth, angle=0]{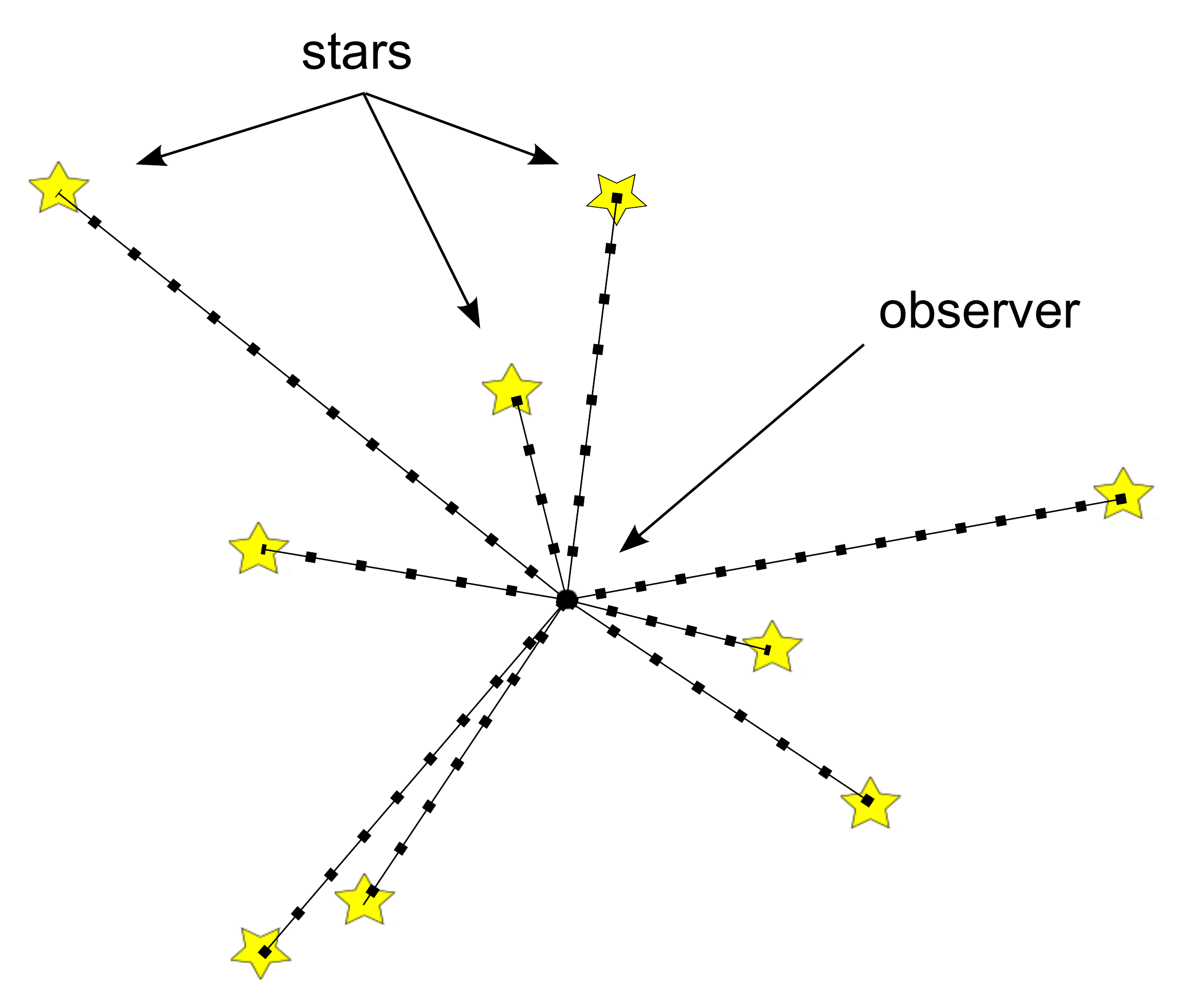}
\includegraphics[width=0.40\textwidth, angle=0]{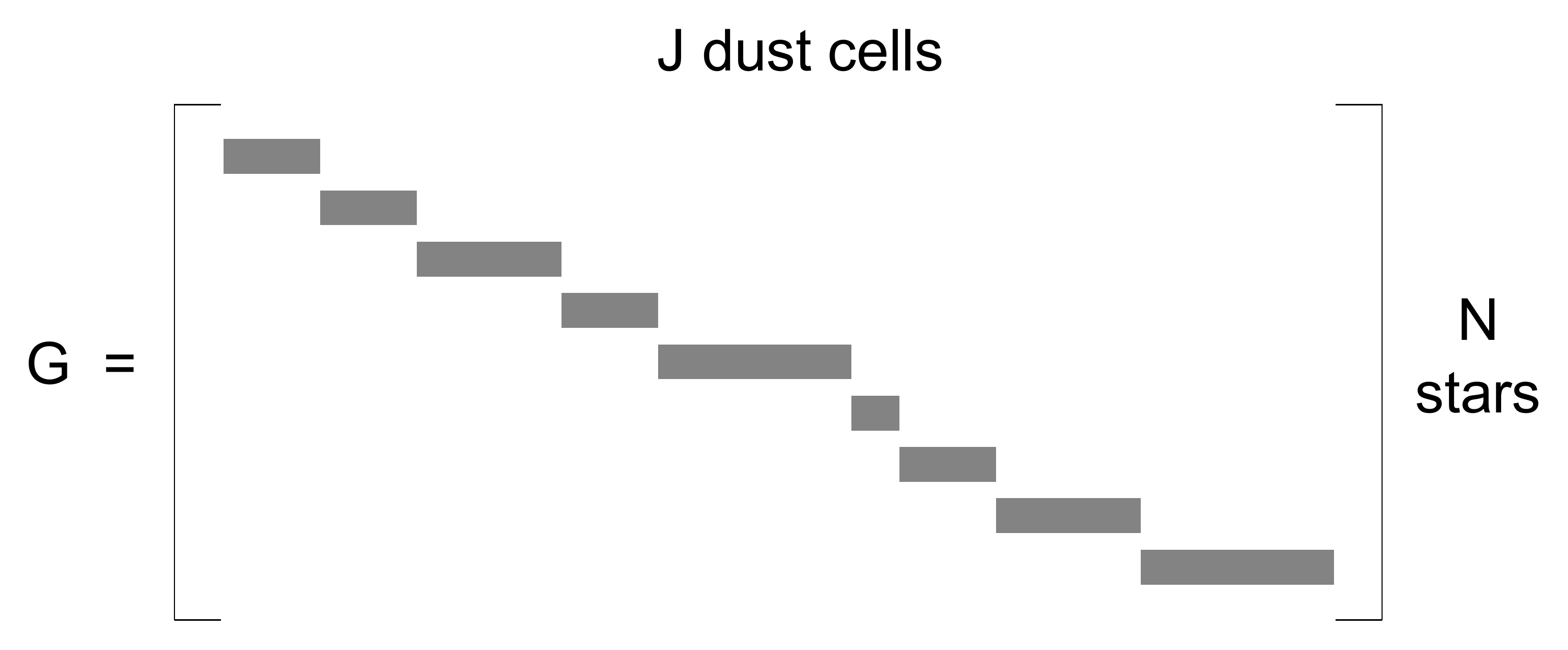}
\caption{Dust cell geometry. Top: The l.o.s towards each of the $N$ stars is divided up into a number of dust cells. The centre of each cell is shown as a small black square. (Here we show cells of constant size.) The total number of dust cells (towards all stars) is $J$, and the length of dust cell $j$ towards star $n$ is denoted $\geonj$. Bottom: These cells are represented by a sparse matrix $\geomat$ of size $N\mby J$. Each row has non-zero elements just for the cells along the l.o.s to that star.}
\label{fig:geometry2}
\end{center}
\end{figure}

Let $\rho(\rvec)$ be the dust density at vector position $\rvec$ measured from the observer.
A model for the attenuation of starlight caused by this dust for a star at position $\rvec_n$ is
\begin{equation}
\estextn \,\propto\, \int_0^{r_n} \rho(\rvec) \, dr
\label{eqn:dustint}
\end{equation}
where $r = | \rvec |$.
The principle of our method is to invert the above to get $\rho(\rvec)$ for an arbitrary point in space given measurements
of the attenuation towards multiple stars. If we adopt a parametric form for $\rho(\rvec)$ then this is straightforward, but the result would be highly limited by the form adopted. Here we use a non-parametric model by dividing the 
pencil beam along the l.o.s toward each star into several dust cells, as shown in figure \ref{fig:geometry2} (top).
Let the (unknown) average dust density in cell $j$ towards star $n$ be $\rhonj$. The integral in equation~\ref{eqn:dustint} can then be replaced by a sum
\begin{equation}
\estextn \,=\, \sum_j \geonj \rhonj \ .
\label{eqn:dustsum}
\end{equation}
The attenuation is unitless. $\geonj$ (the ``geometric factor'') is the length of the cell along the l.o.s. Thus $\rhonj$, which we think of as the ``dust density", has units of attenuation per unit length. As we will use parsecs for distances, $\rhonj$ has units pc$^{-1}$. Let $\extn$ be a measurement of the attenuation towards star $n$. Adopting a Gaussian noise model with standard deviation $\extsdn$ means that the probability of the measurements is
\begin{equation}
P(\extn | \{\rhonj\}) \,=\, \frac{1}{\sqrt{2\pi}\extsdn} \exp\left[ -\frac{1}{2\extsdn^2} (\extn - \estextn)^2 \right] 
\label{eqn:likeA}
\end{equation}
where $\{\rhonj\}$ denotes just those cells on the l.o.s towards star $n$.
If the non-extincted source intensity is $I_0$, then we assume that the observed intensity due to an attenuation $\extn$ is
\begin{alignat}{2}
I \,=\, I_{\circ} \, e^{-\extn} .\
\label{newa}
\end{alignat}
This is related to the measured extinction $A_n$ (in magnitudes) via the usual expression
\begin{alignat}{2}
A_n \,=\, -2.5 \log_{10} \left(\frac{I}{I_{\circ}}\right) \
\end{alignat}
which gives $A_n \simeq 1.0857 \extn$.

Let us suppose that we measure the extinction toward $N$ stars and use a total of $J$ dust cells, for all stars. We define $\geomat$ as the $N\mby J$ matrix with elements $\geonj$, such that the $n^{th}$ row of $\geomat$ contains the geometric factors just for star $n$. 
With this particular geometry
this matrix is very sparse (see figure \ref{fig:geometry2}), because
most row elements are zero (corresponding to the cells for all other stars), and each column has just one non-zero element (stars do not share l.o.s).
Writing the set of dust densities in all cells (for all l.o.s) as the $J$-dimensional vector $\rhovecJ$, 
and the model predictions for the attenuation towards the $N$ stars as $\estextvecN$, 
we can write equation~\ref{eqn:dustsum} as
\begin{equation}
\estextvecN \,=\, \geomat \rhovecJ \ .
\label{eqn:dustsumvec}
\end{equation}
Writing the $N$ attenuation measurements as the vector $\extvecN$ with covariance 
$\likecovN$, then we can generalize equation \ref{eqn:likeA} to be an 
$N$-dimensional Gaussian 
\begin{alignat}{2}
P(\extvecN | \rhovecJ) \,&=\, \frac{1}{(2\pi)^{N/2}|\likecovN|^{1/2}}  \nonumber\\
   \,&\, \exp\left[ -\frac{1}{2} (\extvecN - \geomat\rhovecJ)\trans \likecovN\inv (\extvecN - \geomat\rhovecJ) \right]  \ . 
\label{eqn:likeN}
\end{alignat}
The above equation is the likelihood: the probability of the data given the model parameters.

Our goal is to estimate the dust density $\rhoJp$ at an arbitrary point $\rvec_{J+1}$ in 3D space, given $N$ measurements of the attenuation, $\extvecN$, at known positions.  Put probabilistically, we want to find $P(\rhoJp | \extvecN)$.

It should be noted that each element of $\rhovecJ$ refers to the average dust density in the corresponding cell, although we can consider it to be the dust density at the centre of the cell.  $\rhoJp$, in contrast, is the density at the {\em point} $\rvec_{J+1}$. There is no concept of a cell for points where we want to predict the dust density.

We have $J\geq N$ and $N\gg1$.  A typical problem may involve $N=10^4$ and $J=10^5$ (of the order of ten cells per star on average). In order to infer the dust densities, we need to introduce some connection between the l.o.s, otherwise we just have $N$ independent equations like equation~\ref{eqn:dustsum} with $J$ unknowns, which would be insoluble.

\subsection{Gaussian process model}\label{GP}

We connect the l.o.s using a Gaussian process \citep[e.g.][]{gibbs1997efficient,rasmussen2006gaussian}. This states that the joint probability distribution of the dust density at any $J$ different points (or cell centres) is a $J$-dimensional Gaussian, with a covariance matrix, $\gpcovJ$, which depends on the distance between the points (or cell centres), i.e.\
\begin{equation}
P(\rhovecJ) \,=\, \frac{1}{(2\pi)^{J/2}|\gpcovJ|^{1/2}} \exp \left[ -\frac{1}{2} \rhovecJ\trans \gpcovJ\inv \rhovecJ \right] \ .
\label{eqn:gpJ}
\end{equation}
We assume a zero mean Gaussian in order to have zero values for the dust density in regions where we don't have any constraints from the data.
A useful property of Gaussian processes is that the conditional distribution, $P(\rhoJp | \rhovecJ)$, is also Gaussian. A Gaussian process is just a way of specifying a prior on the covariances between points, as opposed to specifying the functional form of the dust variation in physical space (which is what a parametric model would usually do). This permits a much wider form of functional variations than a parametric model.

\begin{figure} 
\begin{center}
\includegraphics[width=0.30\textwidth, angle=0]{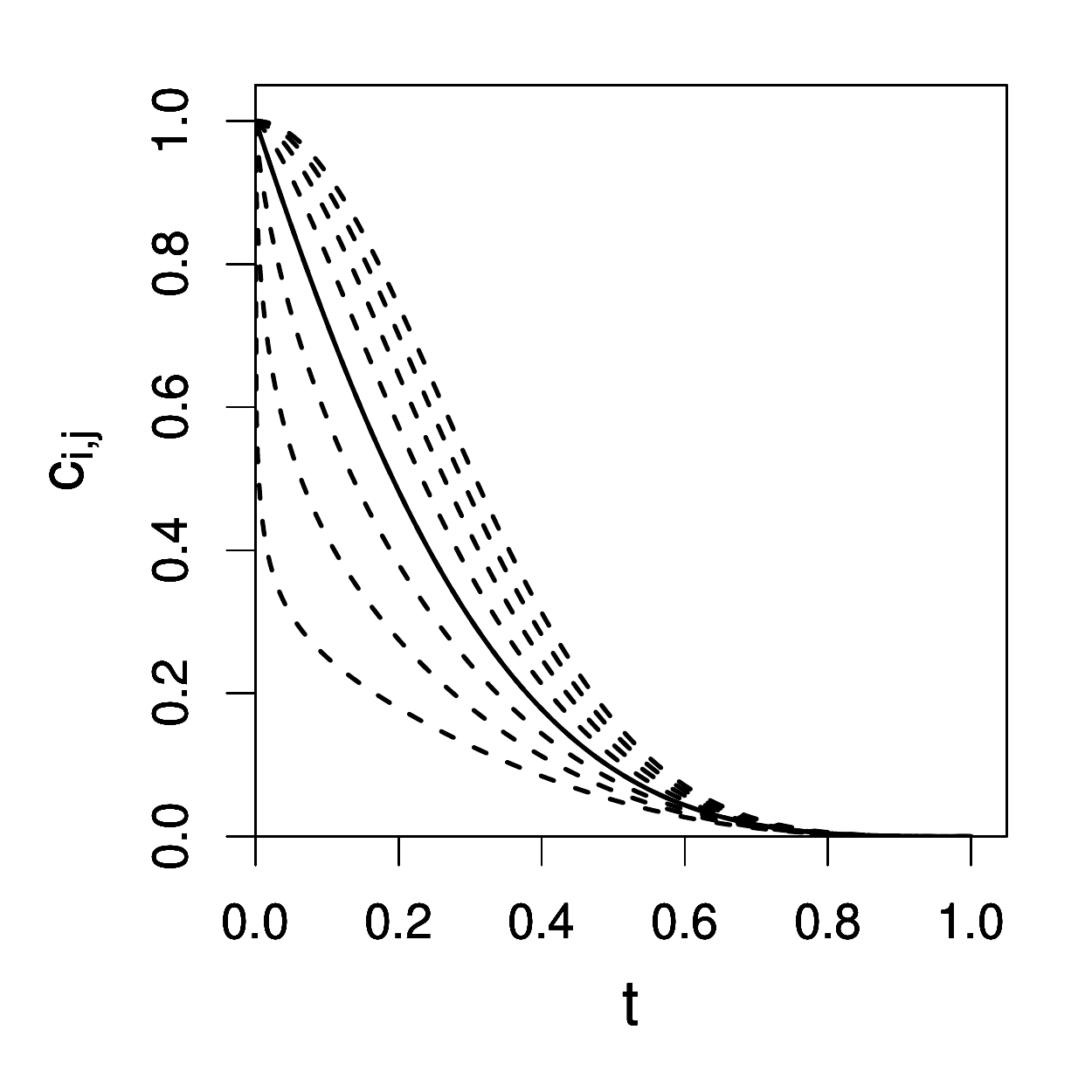}
\caption{Covariance function in equation \ref{eqn:covfunc}. The solid line is the case of $\alpha = 1$ (used in this work) and the dashed lines, from bottom to top, are for different values of $\alpha$ from 0.25, 0.5, 0.75, 1.25, 1.5, 1.75, to 2.}
\label{fig:gneiting}
\end{center}
\end{figure}

An important aspect of the Gaussian processes is to choose an appropriate covariance function. This determines the elements, $\gpcovel_{i,j}$, of the covariance matrix between two points (or cells) $i$ and $j$, with position vectors
$\rvec_i$ and $\rvec_j$, respectively. Here we use a covariance function from \citet{gneiting2002compactly}
\begin{equation}
 \gpcovel_{i,j} = \left\{\begin{array}{lll}
             \theta\, {(1+{t}^{\alpha})}^{-3} \left[ (1-t)\cos({\pi}t) + \frac{1}{\pi}\sin({\pi}t) \right] && {\rm if}~~0 \leq t \leq 1\\
             0 &  & {\rm otherwise} \\
            \end{array}\right.
\label{eqn:covfunc}
\end{equation} 
where 
\begin{alignat}{2}
t \,=\, \frac{|\rvec_i - \rvec_j|}{\lambda} \hspace*{1em} {\rm and} \hspace*{1em} \theta > 0 \: {\rm ,} \: \lambda > 0 \ .
\nonumber
\end{alignat} 
which we illustrate in figure \ref{fig:gneiting}. We use $\alpha = 1$ (the solid line). A larger value of $\alpha$ (e.g. $\alpha = 2$, which has zero gradient at zero separation) produces functions with a smoother spatial variation (plots for $\alpha = 1$ are shown later).
The covariance drops monotonically as the separation between the points increases, to a value of zero once the points are separated by more than $\lambda$ (the scale length). 
Note, however, that the covariance already drops to half its maximum at $t=0.2$, so the effective correlation distance is much less than $\lambda$.
The hyperparameter, $\theta$, in the covariance function determines the overall scale of variations of the dust density. Here we consider the two hyperparameters, ${\lambda}$ and ${\theta}$, to be fixed (see section \ref{simulations}), although they can be inferred from the data.

This specific form of the covariance function does not have a particular physical motivation, but it has compact support, which means it goes exactly to zero beyond some distance: two points separated by more than $\lambda$ will not influence each other. While other covariance functions such as an exponential will give negligible covariance beyond many scale lengths, a truncated covariance function has the advantage of giving rise to sparse covariance matrices, which reduces memory use and accelerates computations. It is important to note that although one can easily write down any function which is truncated, this is not sufficient for it to be a \emph{covariance} function \citep[see][]{rasmussen2006gaussian}.

We can now use this Gaussian process prior together with the likelihood (equation \ref{eqn:likeN}) to determine
$P(\rhoJp | \extvecN)$. This will result in estimating $J\mplus 1$ parameters from $N$ measurements. As $J\geq N$, this means that the resulting density estimates will not be independent. 
This is of course the whole point: to introduce correlations between the dust cells to make the problem tractable and -- more significantly -- to allow us to infer a PDF over the dust density at unobserved points.

It should be noted that the Gaussian model allows the dust density, $\rho$, to be negative. Although this is unphysical, it can be tolerated and will be discussed later on. 
\begin{figure*}
\resizebox{\hsize}{!}{\includegraphics[clip=true]{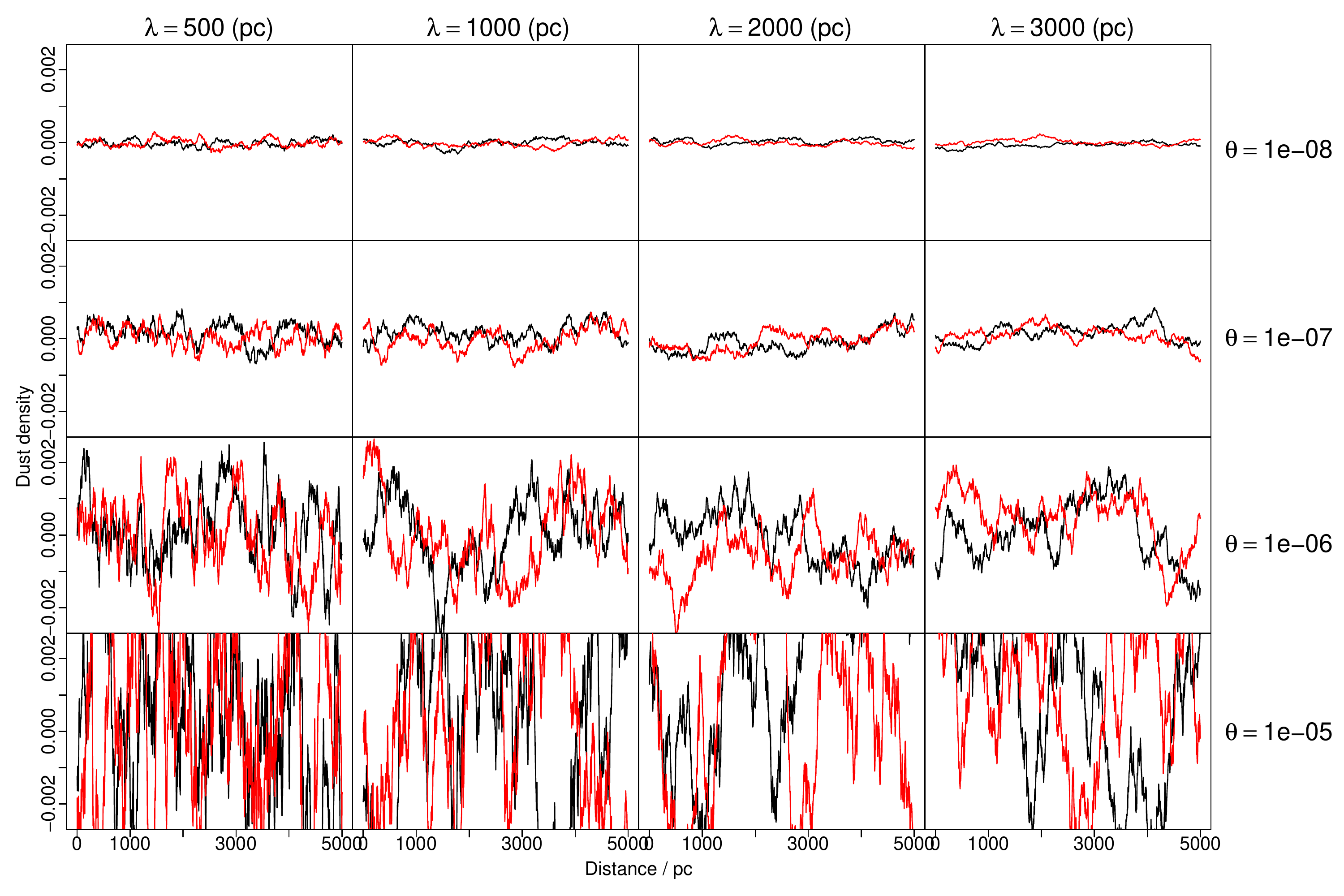}}
\caption{Dust density (attenuation per parsec) drawn from the prior with different hyperparameters of the covariance function (equation \ref{eqn:covfunc}). We define a 1D grid of 1000 equally-spaced points from $ r=0$ to $r=5000 pc$ and each red and black line shows one draw from a 1000-dimensional Gaussian for a fixed $\lambda$ and $\theta$.}
\label{fig:mosaicpriordraws}
\end{figure*}

Figure \ref{fig:mosaicpriordraws} shows samples drawn from the prior for different values of the hyperparameters. To make this we define a 1D grid of 1000 equally-spaced points from $ r=0$ to $ r=5000 pc$. One draw from this 1000-dimensional Gaussian gives us 1000 points which are plotted at their respective positions in space, and then connected with a line. Here, for each fixed pair of $\lambda$ and $\theta$, we draw two samples from the prior (red and black lines). The sharpness of the function is due to the shape of our covariance function (fig. \ref{fig:gneiting}) which allows for sharper fluctuations. As mentioned earlier, a larger value of $\alpha$ in equation \ref{eqn:covfunc} produces smoother variations. It is clear that higher values of $\theta$ result in sharper and larger amplitude variations. Larger values of $\lambda$ produce smoother variations. As we are using a zero-mean prior, we do see negative values for the dust density in the prior draws. This is of course unphysical, but we will see that in the presence of good data our posterior is mostly determined by the likelihood rather than the prior. We come back to this point in section \ref{discussion}.

\subsection{Analytic solution}
Using the law of marginalization over each l.o.s towards observed stars and then applying Bayes theorem, we can write the posterior PDF of the dust density at a given point given the data as
\begin{alignat}{2}
P(\rhoJp | \extvecN) \,&=\, \int P(\rhoJp, \rhovecJ | \extvecN) \, d\rhovecJ \nonumber\\
   \,&=\, \int \frac{P(\rhoJp, \rhovecJ) P(\extvecN | \rhoJp, \rhovecJ)}{P(\extvecN)} \, d\rhovecJ \nonumber\\
   \,&=\, \frac{1}{P(\extvecN)} \int P(\rhoJp, \rhovecJ) P(\extvecN| \rhovecJ) \, d\rhovecJ \ 
\label{eqn:rhointA}
\end{alignat}
where in the last line we use the fact that $\extvecN$ is independent of $\rhoJp$ once conditioned on $\rhovecJ$.
This is a $J$-dimensional integral evaluated over all values of each component of $\rhovecJ$.
The term outside the integral is independent of $\rhoJp$ so is just part of the normalization constant.
The first term under the integral is the Gaussian process prior (equation \ref{eqn:gpJ}), now in $J\mplus 1$ dimensions. The second term is the likelihood (equation \ref{eqn:likeN}). Both are Gaussians, but not in $\rhoJp$. Yet because their arguments 
are linear functions of $\rhoJp$, the integral will have an analytic solution.
Let
\begin{equation}
\rhovecJp \,=\, \left[ \begin{array}{c} \rhovecJ \\ \rhoJp \end{array} \right] 
\end{equation}
be the concatenation of the $J$ dust densities with the dust density at the new point.
Denote its covariance matrix as $\gpcovJp$. The distribution of $\rhovecJp$ follows equation \ref{eqn:gpJ} with $J \rightarrow J\mplus 1$. We partition the inverse of this covariance matrix by writing it as 
\begin{equation}
\gpcovJp\inv \,=\, \begin{bmatrix} \mmatJ & \mvecJ \\ \mvecJ\trans & \mu \end{bmatrix} 
\end{equation}
where $\mmatJ$ is a $J\mby J$ matrix, $\mvecJ$ is a $J\mby 1$ vector, and $\mu$ is a scalar.
We show in appendix \ref{sec:analytic_solution} that the result of the integration is a Gaussian with 
mean $-\beta/\alpha$ and variance $1/\alpha$
\begin{equation}
P(\rhoJp | \extvecN) \,=\, \sqrt{\frac{\alpha}{2\pi}} \exp \left[   -\frac{\alpha}{2}\left(\rhoJp + \frac{\beta}{\alpha}\right)^2 \right] \ 
\label{eqn:rhointF}
\end{equation}
where
\begin{alignat}{2}
\alpha \,&=\, \mu - \mvecJ\trans\rmatJ\inv\mvecJ \nonumber \\
\beta \,&=\, \extvecN\trans\likecovN\inv\geomat\rmatJ\inv\mvecJ \ \nonumber \\
\rmatJ \,&=\, \mmatJ + \geomat\trans\likecovN\inv\geomat \ .
\end{alignat}
In order to calculate the one-dimensional PDF over any single $\rhoJp$ at position $\rvec_{J+1}$, we must 
invert several matrices of size $J$, and this takes time $\ofo(J^n)$, where $n \lesssim 3$ for exact matrix inversion. The calculations can, however, be accelerated using certain matrix identities, as described in appendix \ref{sec:acceleration}.

\section{Demonstration using simulated data}\label{simulations}

In this section, we present some results based on both simple mock data and simulated data from Gaia Universal Model Snapshot (GUMS) \citep{2012A&A...543A.100R}. We use these results to explain the main properties and behaviour of the model.

\subsection{Simple mock data}\label{simulated_data}

We first investigate the ability of our model to capture structures in the dust distribution and to infer it in unobserved regions. We further investigate the influence of the two model hyperparameters, $\theta$ and $\lambda$, on the dust density models produced.
\begin{figure} 
\begin{center}
\hspace*{-2em}\includegraphics[width=0.45\textwidth, angle=0]{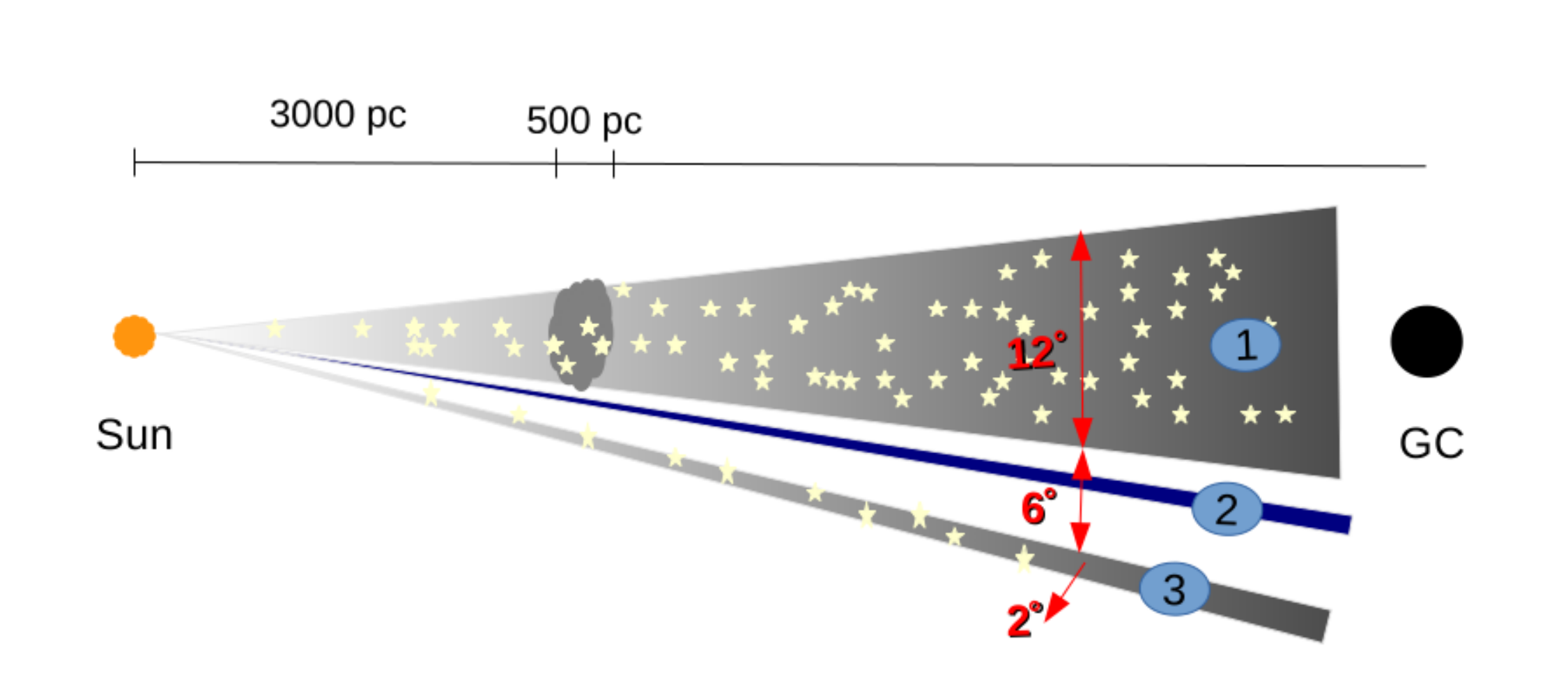}
\caption{Illustration of the simple mock data in the Galactic plane.
The true dust density decreases exponentially from the Galactic center (GC) in all directions. The observer is at the Sun. 
Region 1: in addition to the main dust variations there is a cloud between 3 and 3.5 kpc. 200 stars are observed here. Region 2: no stars are observed. Region 3: 100 stars are observed here (there is no cloud). Stars are drawn from throughout regions 1 and 3, their attenuations calculated, and observational noise added.}
\label{fig:simulationcartoon}
\end{center}
\end{figure}
\begin{figure} 
\begin{center}
\includegraphics[width=0.50\textwidth, angle=0]{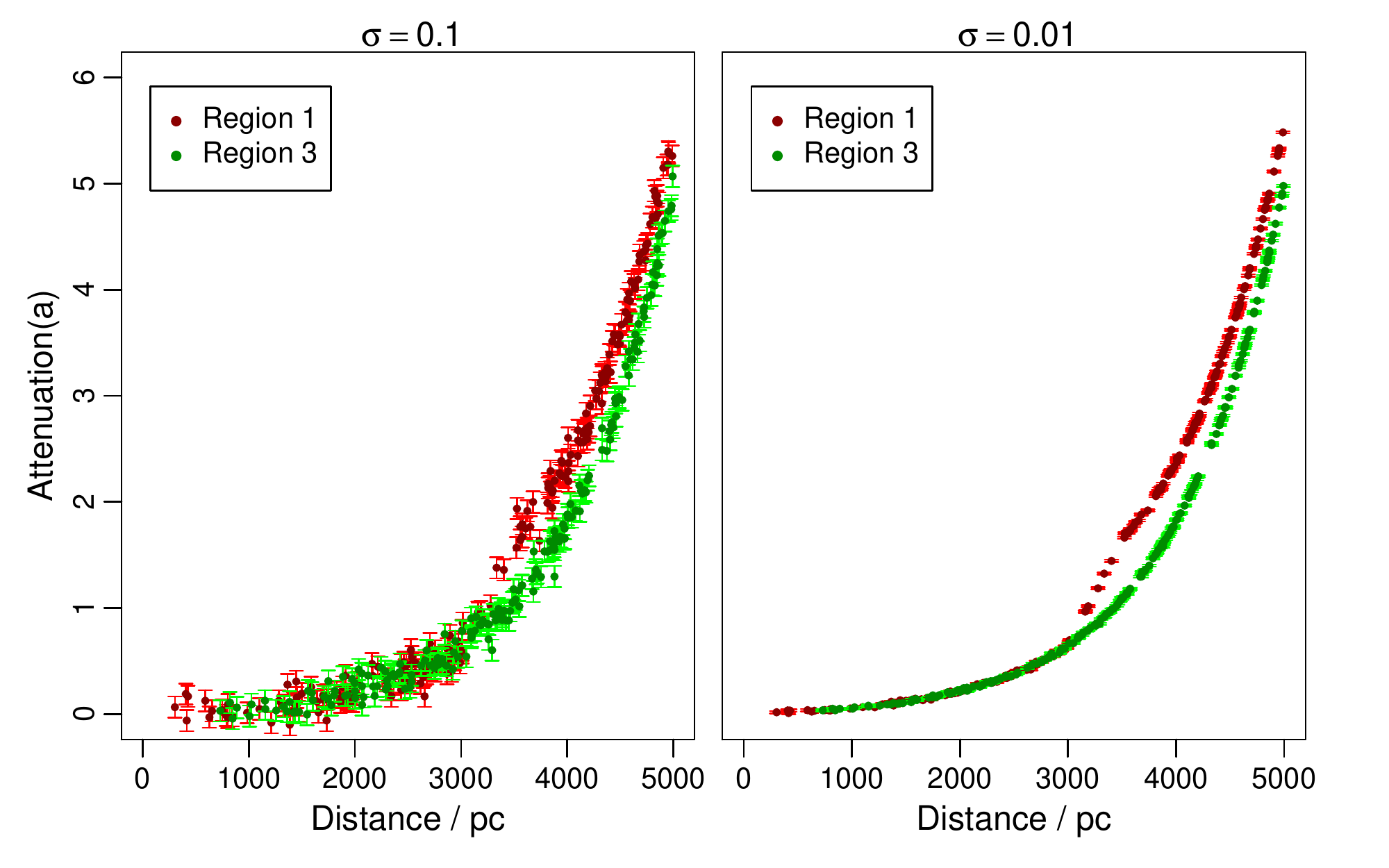}
\caption{Attenuation in the simple mock data set as a function of distance for two different attenuation uncertainties ($\sigma_n$ in equation \ref{eqn:likeA}) of 0.1 (left) and 0.01 (right) (no units).
An increase in attenuation where the cloud is located (3 kpc to 3.5 kpc) is evident at the smaller noise level (right panel). It should be noted that attenuation is unitless (see equation \ref{eqn:dustsum}).}
\label{fig:simulationextinction}
\end{center}
\end{figure}

Figure \ref{fig:simulationcartoon} illustrates the simulation set up, which specifies the distribution of dust and observations of stars in the Galactic plane.  There is a general distribution of dust, the density of which decreases exponentially from the Galactic Center in all directions with a length scale of 1 kpc. Observations are made in region 1 (a 12$^{\circ}$ wedge between longitudes 354$^{\circ}$ and 6$^{\circ}$) and region 3 (a 2$^{\circ}$ wedge between longitudes 12$^{\circ}$ and 14$^{\circ}$), but not in region 2, a narrow wedge ($l=7^{\circ}-8^{\circ}$) which lies between them. In region 1 there is an additional dust cloud of 500 pc depth centered on a distance of 3.25 kpc from the Sun. This does not extend into region 3 (whether it extends into region 2 in ``reality'' is immaterial as we have no observations there).  We want to infer the distribution of the dust density over all three regions using measurements of the attenuations towards 200 stars spread uniformly over region 1, and of 100 stars spread uniformly over region 3.  Two different sets of noisy measurements are considered: standard deviations ($\sigma_n$ in equation \ref{eqn:likeA}) of 0.1 and 0.01 on the attenuations.  Figure \ref{fig:simulationextinction} shows the noisy input data for these two situations. In region 3, the dust density increases roughly exponentially with increasing distance from the Sun. The attenuation is the integral of this, so is more or less exponential too. Region 1 is similar, but has the dust cloud in addition. This can be made out reasonably well in the data at higher signal-to-noise (right panel), but is barely noticeable at the lower signal-to-noise (left panel).

\begin{figure*}
\begin{center}
\includegraphics[width=0.70\textwidth, angle=0]{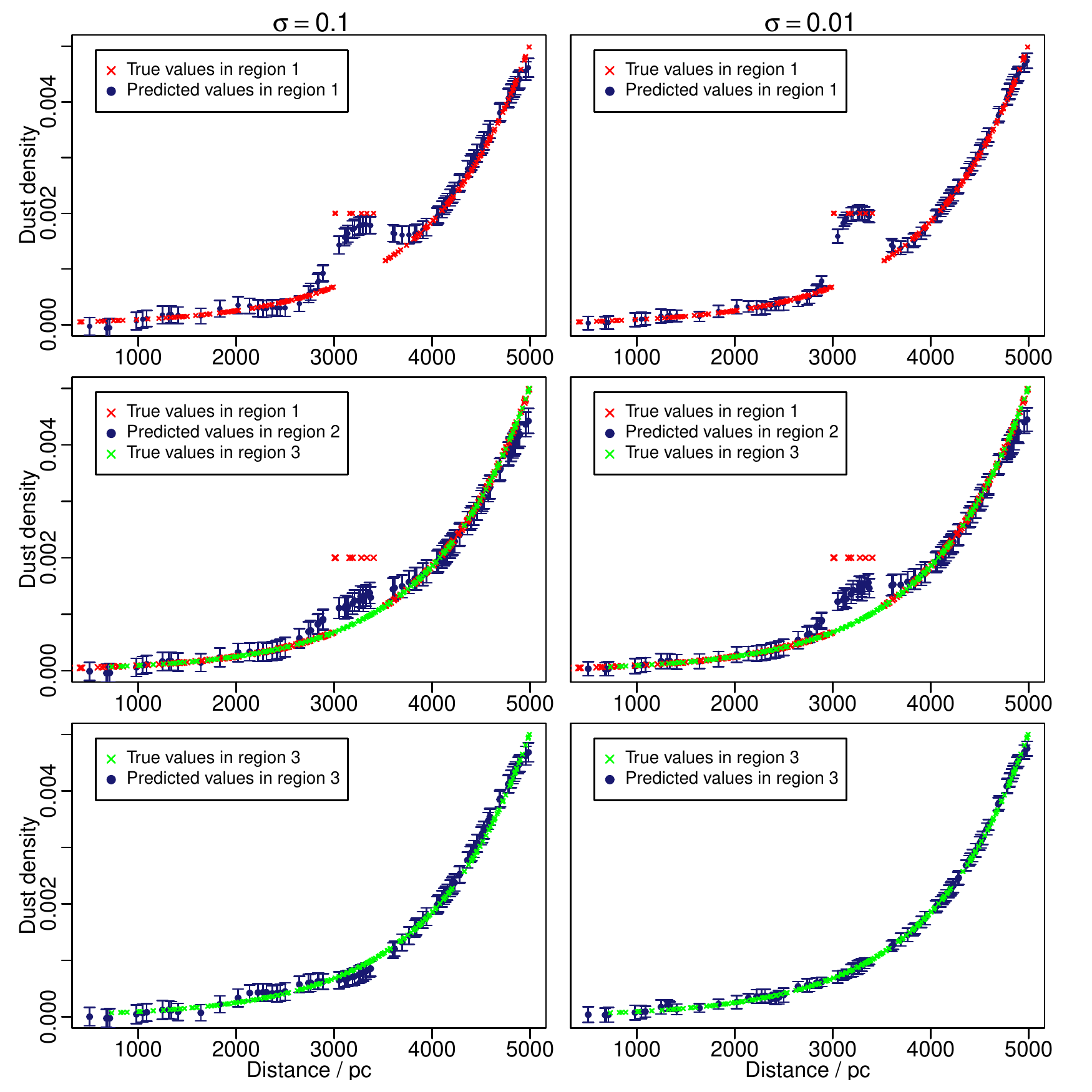}
\caption{Predicted dust densities (attenuation per parsec) in the different regions of figure \ref{fig:simulationcartoon} and for two different attenuation errors: 0.1 (left) and 0.01 (right). Red and green crosses show true values for regions 1 and 3 respectively, and blue points show predicted values for different regions (1, 2 and 3 from top to bottom). 
The error bars on the latter points are also predicted by our model.
Although there are no observations in region 2 the model can still predict dust densities there.}
\label{fig:mosaicpred}
\end{center}
\end{figure*}
We use our model 
to estimate the dust density at 100 points distributed at random at each of the three regions.
We set the hyperparameters to $\theta=10^{-7}$ and $\lambda=2$\,kpc (equation \ref{eqn:covfunc}) 
and use uniform cell sizes of size $g=250$\,pc. The choice of these values will be discussed later in this section.
Figure \ref{fig:mosaicpred} shows (as blue points) the estimated dust densities as a function of distance, as well as the uncertainty on this estimate (as error bars): these are the mean and standard deviation of the Gaussian posterior in equation \ref{eqn:rhointF}. These predictions can be
compared to the true values for regions 1 and 3, which are shown as red and green crosses respectively.
We see that the inference of the overall exponentially-varying dust is good in all three regions, including in region 2, where there were no observations. This shows that our model performs sensible, plausible interpolations across unobserved regions. This is possible because of the smoothness prior imposed by the Gaussian prior.
For region 1 (top row), the model predicts the location and density of the dust cloud well, even at the lower signal-to-noise ratio. In that case the estimated uncertainties (error bars) are also larger, which is what we want from a model.
In region 2 (middle row), we ask the model to predict dust densities along a very narrow wedge between $l=7^{\circ}$ and $l=8^{\circ}$, located between regions 1 and 3 but closer to region 1 (with cloud) than region 3 (without cloud). With $\sigma = 0.01$ (right), an increase is obvious at 3 to 3.5 kpc corresponding to the distance of the dust cloud in region 1. A smaller increase is visible for the larger noise case (left). This too is a sensible interpolation of the available data: the cloud must stop or peter out somewhere between region 1 and region 3 because it is no longer observed in region 3. We have no information on where, but the
covariance prior tells is that 
the closer we are to region 1, the more likely we are to still encounter the cloud.
(It should be noted that the physical transverse extent from $6^{\circ}$ to $12^{\circ}$ at 3\,kpc is much less than the length scale, $\lambda$, we have adopted.)
In region 3, the model predictions show no indication of a cloud: they are influenced primarily by the nearer, cloud-free attenuation estimates.

The length scale, $\lambda$, sets the maximum distance over which dust cells are correlated.  Note, however, that the correlation is only significant for values considerably smaller than $\lambda$ (see figure \ref{fig:gneiting}).  Choosing a value of $\lambda$ that is too small will result in too many cells being disconnected (or having very low correlations), with the outcome that the information in the data is propagated less well.
A value of $\lambda$ that is too large will make even quite distant cells relatively highly correlated, potentially blurring out the local variance determined by the data.

The cell size (here taken as constant) is the radial length over which we assume the dust density to be constant when setting up the model.  It is only used by equation \ref{eqn:dustsum} (generally equation \ref{eqn:dustsumvec}) to discretize the dust density for representing the dust attenuation towards observed stars. It is used neither in the calculation of the covariance nor in the computation of dust density at new points, so contrary to possible expectations it does not represent the minimum scale over which we can compute density variations. It does, however, set some kind of minimum length scale over which we are sensitive to dust variations. Ideally we would use very small cells, but the computation time grows as the third power of the number of cells, so in practice we are limited by computational considerations.

The hyperparameter, $\theta$, sets the scale of the covariance and thus the amplitude of variations in the dust.
For a given $\lambda$ and cell size, a larger value of $\theta$ means we can capture larger variations in the dust.
We see from 
equation \ref{eqn:covfunc} with $t=0$ that $\theta$ is the expected variance in the dust at any point.
An estimate for the value of $\theta$ is therefore the variance in the expected distribution of the dust
density over all cells. We can get an order of magnitude estimate of this before applying the model, by
using the simplifying assumption that towards a given star, $n$, every cell has the same dust density, $\rho_n$, and same variance therein, ${\rm Var}(\rho_n)$. 
Adopting a constant cell size $g$ for a given star, it follows 
from equation \ref{eqn:dustsum}, using $a_n$ as our estimate of $f$, that
\begin{equation}
\rho_n \,=\, \frac{a_n}{j_n g}
\label{eqn:rho_est}
\end{equation}
where $j_n g$ is just the distance to the star. 
Let $\mu_\rho$ be the average of the $\{\rho_n\}$ across all $N$ stars. The weighted variance
of this distribution is
\begin{equation}
\theta \,=\, \frac{\sum_{n=1}^{N} \, w_n (\rho_n - \mu_{\rho_n})^{2}}{\sum_{n=1}^N w_n}
\label{eqn:theta}
\end{equation}
where each weight, $w_n$, can be set equal to the inverse of the variance, ${\rm Var}(\rho_n)$,  in the corresponding value of $\rho_n$.
These can be found by taking variance equation \ref{eqn:dustsum} (with $a_n$ as our estimate of $f$) in which $\rho_{n,j} = \rho_n$ is consistent with our above assumptions. This gives
\begin{equation}
{\rm Var}(\rho_n) \,=\, \frac{{\rm Var}(a_n)}{j_n \, g_n^{2}} 
\label{eqn:rho_variance}
\end{equation}
where $j_n$ is the number of cells towards the star $n$. 
Equations \ref{eqn:rho_est}--\ref{eqn:rho_variance} allow us to estimate $\theta$ using the measured attenuations, $a_n$, and their uncertainties, $\sigma=\sqrt{{\rm Var}(a_n)}$ as well as the distances and adopted cell sizes.
We will see that the larger the value of $\theta$, the larger the error bars on our dust density predictions will be, as we expect.
Using the simple mock data, we get $\theta=1.9\times10^{-7}$ for $\sigma=0.1$ and
$\theta=1.8\times10^{-7}$ for $\sigma=0.01$.
\begin{figure*}
\resizebox{\hsize}{!}{\includegraphics[clip=true]{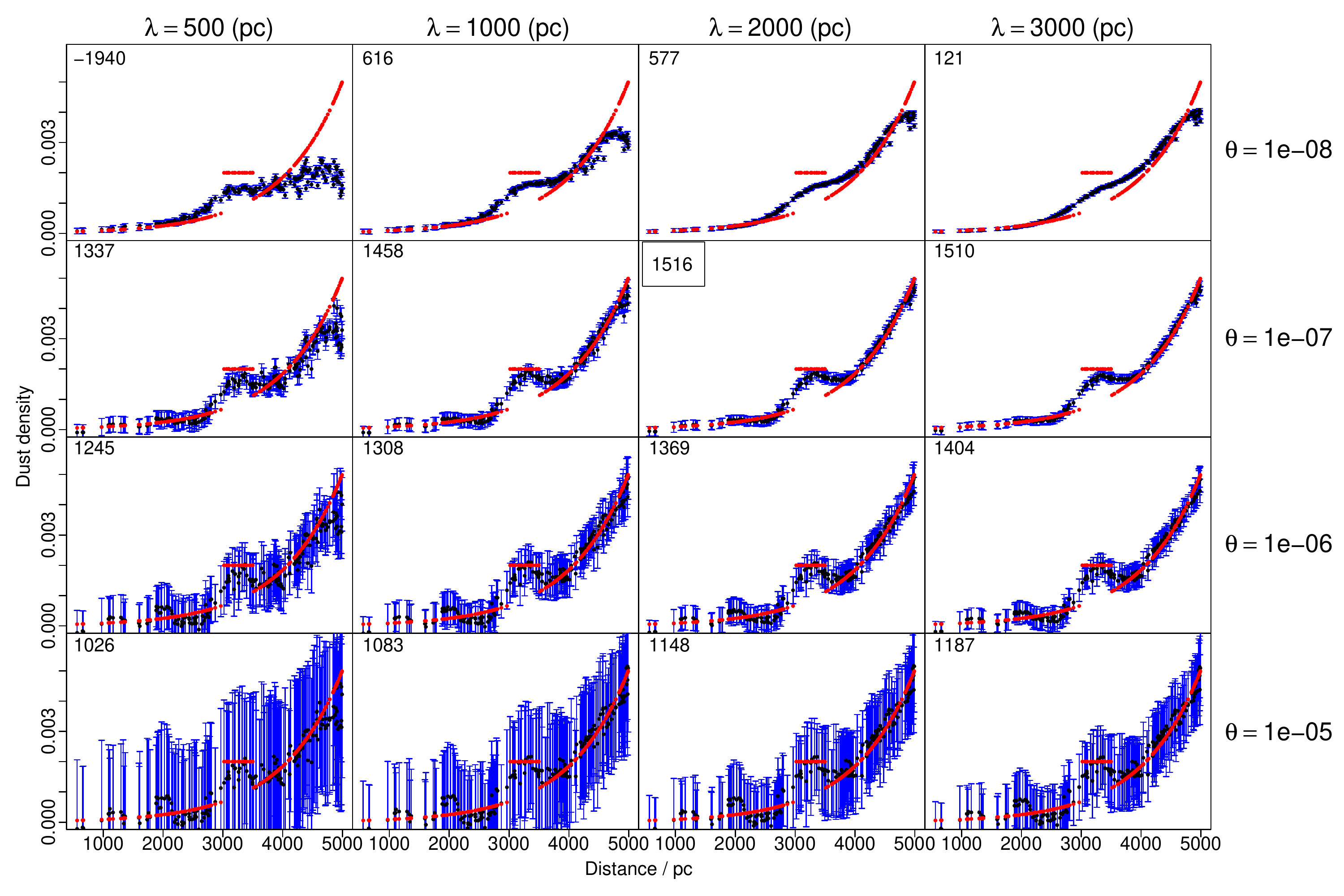}}
\caption{Effects of the hyperparameters on the model predictions (attenuation per parsec) in region 1 with $\sigma=0.1$. Red points are the true values and black points with blue error bars are predictions by the model. 
It is important to note that the predicted uncertainties are highly correlated.
Each column is for different values of $\lambda$ (length scale), and each row is for different values of $\theta$. The number in the top left corner is the natural logarithm of the reconstruction probability, defined by equation \ref{eqn:reconstruction_probability}.
\label{fig:mosaiclambdatheta}}
\end{figure*}

It's important to realize that the minimum length scale over which we can probe dust variations is {\em not} set by $\lambda$. We see in figure \ref{fig:mosaiclambdatheta} and we will see later that we can probe significant variations on length scales much less than $\lambda$. The finite cells sizes ($g$) aside, the minimum length variation is actually set by a combination of $\lambda$ and $\theta$, plus, most importantly, by the data themselves.

Figure \ref{fig:mosaiclambdatheta} shows the effects of varying the values of the hyperparameters on the model predictions.  Here we compare the true and estimated dust density for different values of $\lambda$ (columns) and $\theta$ (rows) for region 1 with $\sigma=0.1$.  As before, red crosses show true values and blue points show the predictions. Values of $\lambda$ and $\theta$ increase from left to right and top to bottom, respectively.  Larger values of $\lambda$ produce smoother variations in the dust. This is largely because this connects more cells, thereby increasing the amount of data used to estimate the dust densities. This also decreases the uncertainties on the estimates (the error bars).  Smaller values of $\theta$ prevent the model from following steep changes in the dust density: for a given $\lambda$ we see smoother variations in the dust density at smaller $\theta$.
 
The very nature of our covariance model is that points separated by less than $\lambda$ have correlated posterior distributions. Thus not only are the predicted dust estimates correlated, but so are their predicted uncertainties.
Thus the error bars of adjacent points in figure \ref{fig:mosaiclambdatheta} are highly correlated. They are simple point estimates of a continuous function (we could make estimate on a grid twice as dense; the error bars would not change).
Significant variations of the dust density can be (and are) obtained which are far smaller than these error bars.

In a real application -- where we don't know the true dust densities -- we could compute the likelihood of the data at the predicted values of the dust densities from equation \ref{eqn:likeN}. We could do this for a range of hyperparameters $\lambda$ and $\theta$ and find the model value which gives the highest likelihood.  This is a form of Bayesian model selection, because the regularizing prior has been used to estimate the individual dust densities.  The different models correspond to different values of the hyperparameters.\footnote{What we call the likelihood here is not the likelihood in the sense of parametric models, which is the probability of the data for a given model {\em and} a given set of parameters. While we could maximize this likelihood to find the best parameters for a given model, we cannot use it to choose among models, because it contains no regularization and so will just identify the most complex model (we can eventually fit the data perfectly, noise and all). The likelihood we are talking about in this Gaussian process context is at a higher level. The ``parameters'', if you will, have effectively been marginalized over by the Gaussian process to produce the probability of the data for given hyperparameters.}  The one practical disadvantage of this is that we would have to compute the dust densities along the l.o.s to every star and for all values of hyperparameters, which could become very time consuming.

For the sake of this simulation -- where we do know the true dust densities -- we compute instead the probability of the inferred dust densities given the true dust densities. By construction, the inferred dust densities have a joint Gaussian distribution, and the true dust densities have no variance. The log probability is therefore given by
\begin{equation}
\ln P = - \frac{1}{2} {\triangle}{\boldsymbol \rho}\trans \gpcov {\triangle} {\boldsymbol \rho} - \frac{1}{2} \ln({(2\pi)}^n|\gpcov|) \ 
\label{eqn:reconstruction_probability}
\end{equation}
which we will call the {\em reconstruction probability}.  It is analogous to the likelihood (or negative log of the sum-of-squared residuals, if the covariance were unity) but with model predicted values replaced by their true ones.  $n$ is the number of predicted points, $\gpcov$ is the $n\times n$ covariance matrix or these points with elements given by equation equation \ref{eqn:covfunc}, and ${\triangle}{\boldsymbol \rho}$ is the $n$ $\times$ 1 vector of the differences between true and predicted dust densities.  The numbers in the corners of the panels in figure \ref{fig:mosaiclambdatheta} show the values of this metric. The highest value is at ${\lambda} = 2000$\,pc and ${\theta} = 1 {\times} 10^{-7}$, which are the values we used for our predictions in figure \ref{fig:mosaicpred}. Recall that our pre-modelling order of magnitude estimate of what to use for $\theta$ gave a value of $\theta=1.9 {\times} 10^{-7}$ for $\sigma=0.1$, which is very similar. This procedure couldn't specify $\lambda$, because we are free to specify this according to the flexibility of the fitting we wish to achieve (and/or our knowledge of the true scale of the variations).  For a fixed $\theta$ of $10^{-7}$ and different values of $\lambda$, we see from figure \ref{fig:mosaiclambdatheta} that as long as $\lambda$ is large enough (a few times cell sizes) to connect many cells in 3D space (${\lambda} =$ 1000, 2000 and 3000\,pc in this example), we can achieve good results. 

From figure \ref{fig:mosaiclambdatheta}, it is clear that the model predicts the dust densities with smaller uncertainties when using larger values of $\lambda$. This is because it then uses more points to predict the dust densities for every new point, although the closer points of course still have more influence (correlation) than more distant points. As the dependence on $\lambda$ is not strong (once $\theta$ is set), we choose to fix $\lambda$ to a relatively large value (a few times the cell size).

\subsection{Gaia Universe Model Snapshot (GUMS)}\label{GUMS}

We now look at a more realistic set of simulations taken from the Gaia Universe Model Snapshot (GUMS) \citep{2012A&A...543A.100R}.

GUMS is a simulation, generated by the DPAC prior to the Gaia launch, of what the Gaia catalogue can be expected to contain. It contains both the intrinsic properties of the objects as well as simulations of the noise-free Gaia observations (or more precisely, the corresponding catalogue products). It comprises around 1.6 $\times$ $10^9$ stars (in single or multiple systems) with G-band magnitudes brighter than 20. We select the coordinates, distance, G-band magnitude, absolute V-band magnitude, $(V - I)$ colour, extinction $A_V$, and effective temperature \teff\ for these
stars. GUMS uses a dust model to generate its extinction values, but this is not part of our simulated catalogue (we have no knowledge of the true dust densities).

To map dust extinction we need not -- and should not -- use all stars. Due to computational limitations, the model, as it stands, cannot cope with anything nearly as large as the number of objects in the catalogue (see section \ref{discussion} for possible improvements). We therefore first select just those stars with $G<15$, parallax errors less than 5\%, and \teff\ between 5\,000 and 10\,000\,K. Using cooler stars, e.g. down to 3\,000\,K, we found that it does not make a significant difference in the results, but as their parallax and/or extinction estimates would often be less precise, we would probably omit them in practice.  The magnitude selection is imposed to ensure that the Gaia spectrophotometry have high signal-to-noise ratio so that $A_V$ is determined to better than about 0.05\,mag by the DPAC processing \citep{2013A&A...559A..74B} and we calculate the expected end-of-mission parallax errors from the simplified formula in \citet{2014EAS....67...23D}.  Although a selection on apparent magnitude biases our sample towards less extinct stars, this is hard to avoid in practice (as almost all surveys have a magnitude limit). As Gaia provides parallaxes from which we can infer distances, we could instead attempt to select all stars within a given volume.  But this selection (magnitude and temperature limit) is better because the distance uncertainties are asymmetric and are themselves a strong function of magnitude, plus the interstellar extinction would significantly limit the size of a complete volume at low Galactic latitudes.  A better approach might be to limit the selection to intrinsically bright stars, over a narrow \teff\ range (and therefore absolute magnitude range) for which we can estimate accurate extinctions. This will be explored in more detail in subsequent work.

\begin{figure} 
\begin{center}
\includegraphics[width=0.50\textwidth, angle=0]{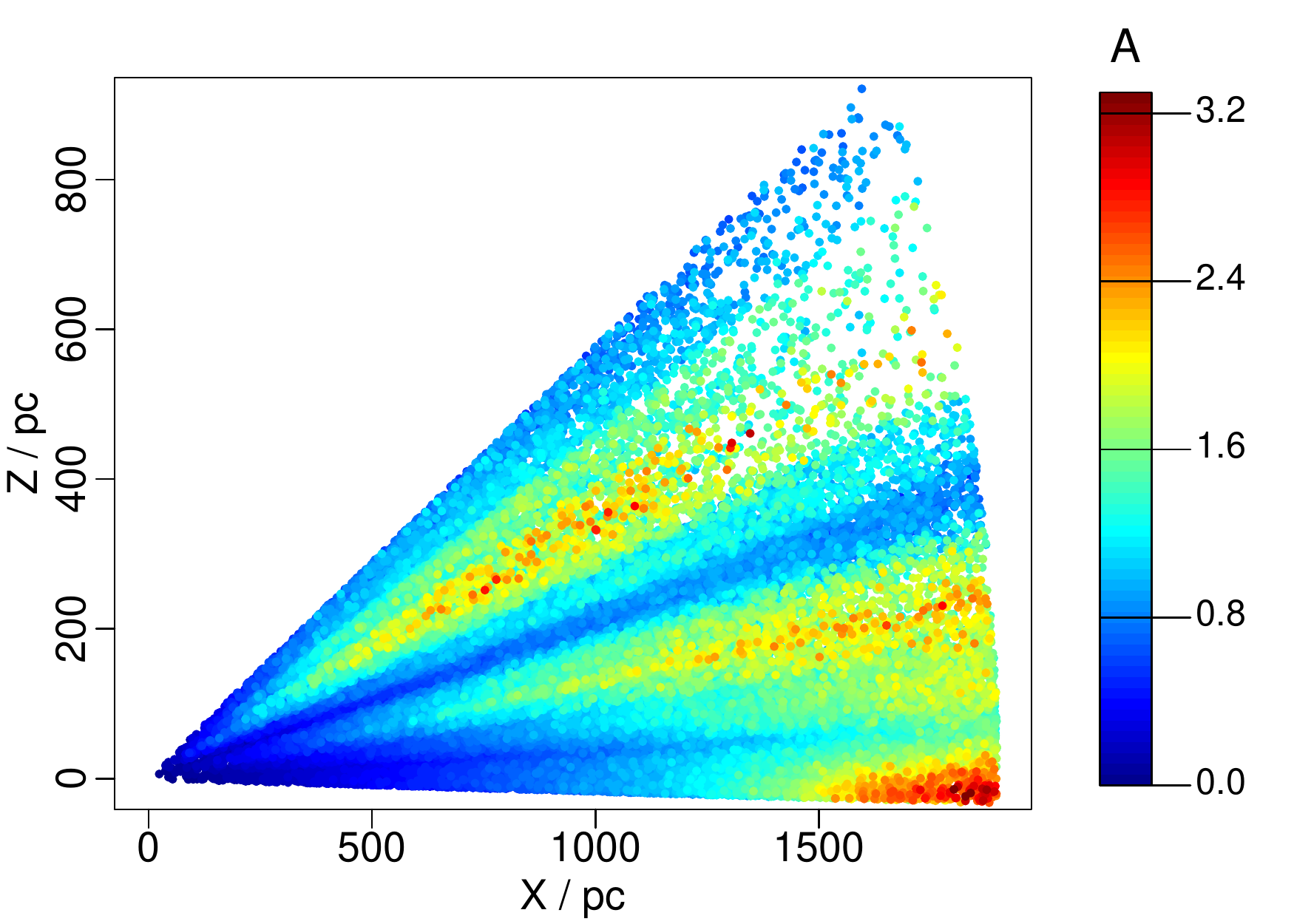}
\caption{Two-dimensional positions of the 52\,000 stars meeting our selection criteria over the region $5^{\circ} < l < 7^{\circ}$ and $-1^{\circ} < b < 30^{\circ}$ out to 2\,kpc from the GUMS catalogue, colour-coded by the true (GUMS) extinctions in magnitudes.}
\label{fig:data_5000}
\end{center}
\end{figure}

Applying the above selections gives us 30 million stars distributed throughout the simulated Galaxy.  To ease the interpretation of the results for the sake of this demonstration, we select stars in a wedge narrow in longitude (5$^{\circ}$ to 7$^{\circ}$) but broad in latitude (-1$^{\circ}$ to 30$^{\circ}$) within 2\,kpc containing around 52\,000 stars.  Their positions and extinctions are shown in figure \ref{fig:data_5000}. The drop off at high latitudes is due to the decline in density of the Galactic disk population away from the plane. The presence of small, nearby regions of higher dust density are apparent from the higher attenuations along some l.o.s.
\begin{figure} 
\begin{center}
\includegraphics[width=0.50\textwidth, angle=0]{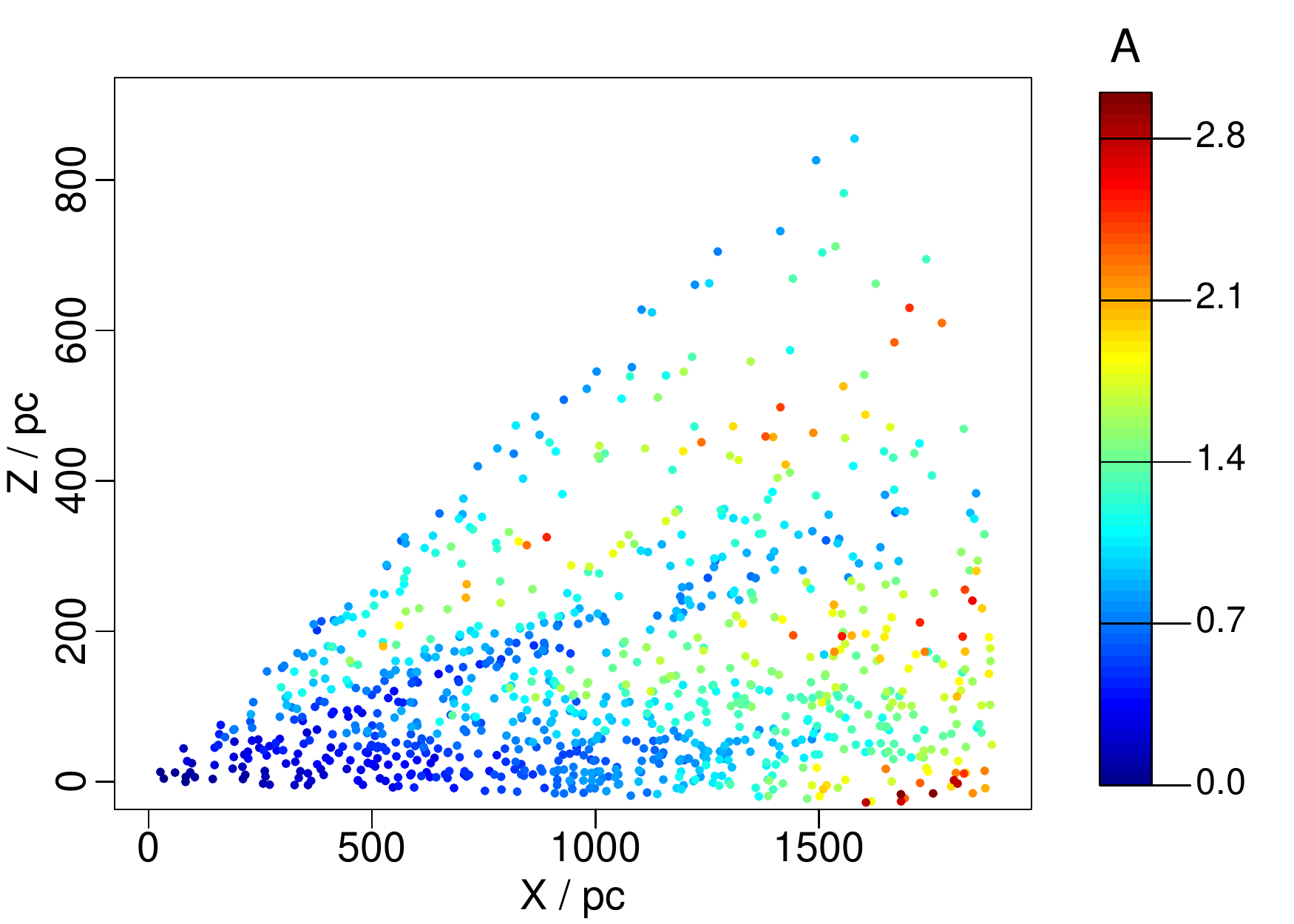}
\caption{Random set of 1000 stars selected from those shown in figure \ref{fig:data_5000}. See section \ref{GUMS}.}
\label{fig:sample_1000_lb}
\end{center}
\end{figure}

A useful feature of our method is that we do not have to use all stars within a region.  Although the stars are independent probes of their l.o.s extinctions, stars in close proximity to one another probe much of the same dust.  Provided the stars retained have a high enough spatial density to map out the minimum scale of the spatial variations we want to probe, randomly removing additional stars will not qualitatively change the resulting map.  It may reduce the precision, because we then have fewer measurements to determine the dust density, but often other factors dominate the uncertainties (we will examine this in section \ref{discussion}). However, as reducing the number of stars can lead to great computational savings and increased numerical stability, this is a useful strategy to pursue.
Figure \ref{fig:sample_1000_lb} shows the positions of 1000 stars randomly selected from the sample of 52\,000.
Although there are far fewer stars, we can still see most of the structures from the full sample.

\begin{figure} 
\begin{center}
\includegraphics[width=0.50\textwidth, angle=0]{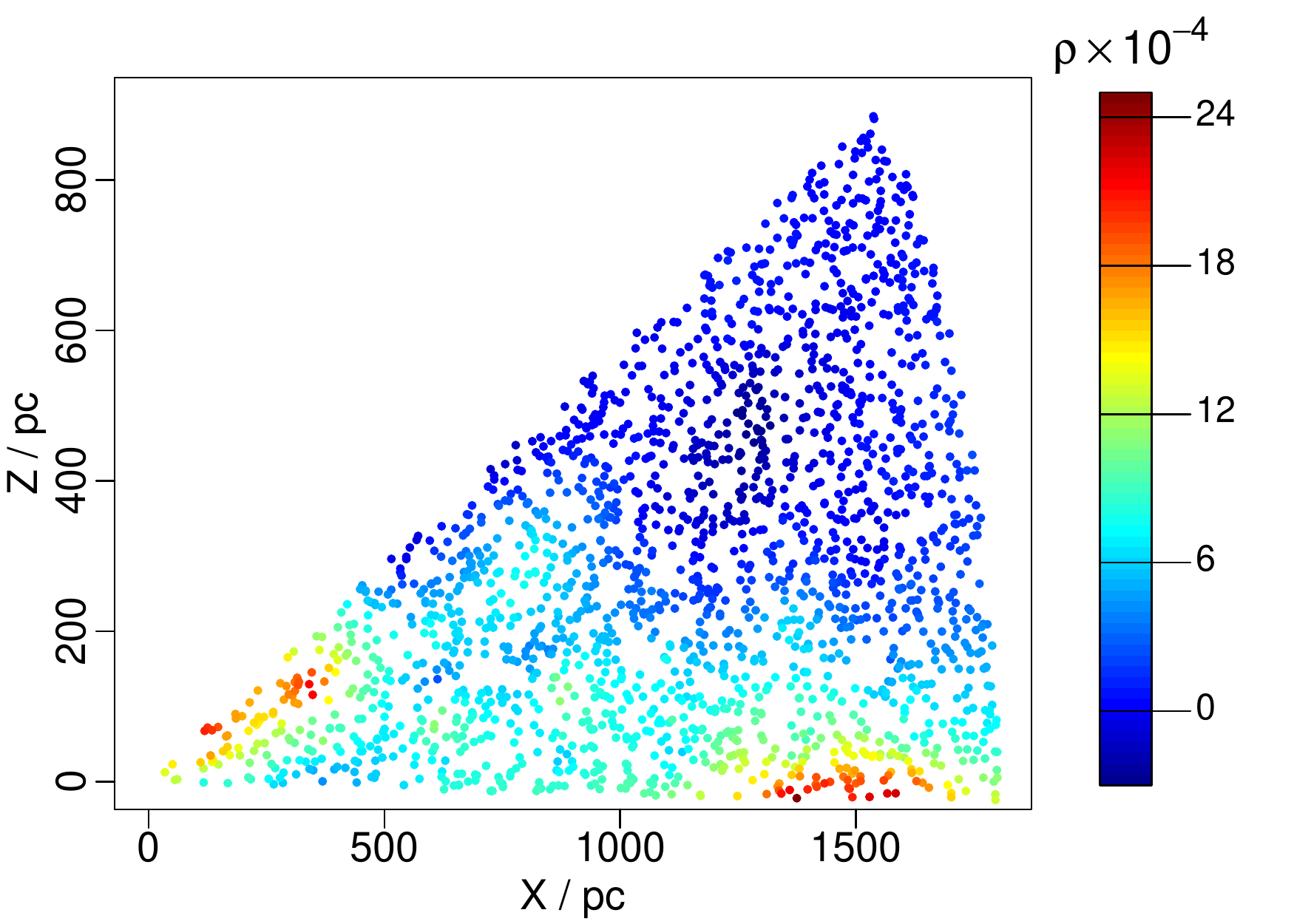}
\caption{Predictions of the dust densities for the selected area from GUMS catalogue ($5^{\circ} {\textless} l {\textless} 7^{\circ}$ and $-1^{\circ} {\textless} b {\textless} 30^{\circ}$) for 2000 new points in two dimension coloured coded by the dust density values. A high value dust cloud is seen at the large distances and low latitudes as well as some lower-value dust clouds in other locations.}
\label{fig:pred_sample1_1000_posl}
\end{center}
\end{figure}

Figure \ref{fig:pred_sample1_1000_posl} shows the predicted dust densities for 2000 random new points in the selected region. A localized region of high dust density (``dust cloud'') 
is apparent at around $(X, Z) = (400, 150)$ which is responsible for the high extinctions beyond this point visible in figure \ref{fig:sample_1000_lb}, namely the upper diagonal wedge.
Likewise, the very high extinction region in the plane beyond about 1400\,pc is assigned a cloud at the same distance.
Less apparent is a dense dust cloud responsible for the intermediate diagonal wedge in figure \ref{fig:sample_1000_lb}; the model has instead attributed this to a more diffuse region of higher dust density, perhaps because there is a larger region of higher extinctions at intermediate latitudes.
It should be noted that the position of these new points have been chosen at random; we are free to select them and predict dust density at any point in 3D space. For this predictions we used a dust correlation length scale, $\lambda$, of 2 kpc, dust cell lengths, $g$, of 250 pc and a dust variation scale, $\theta$, of $1 {\times} {10}^{-7}$.
\begin{figure}
\begin{center}
\includegraphics[width=0.50\textwidth, angle=0]{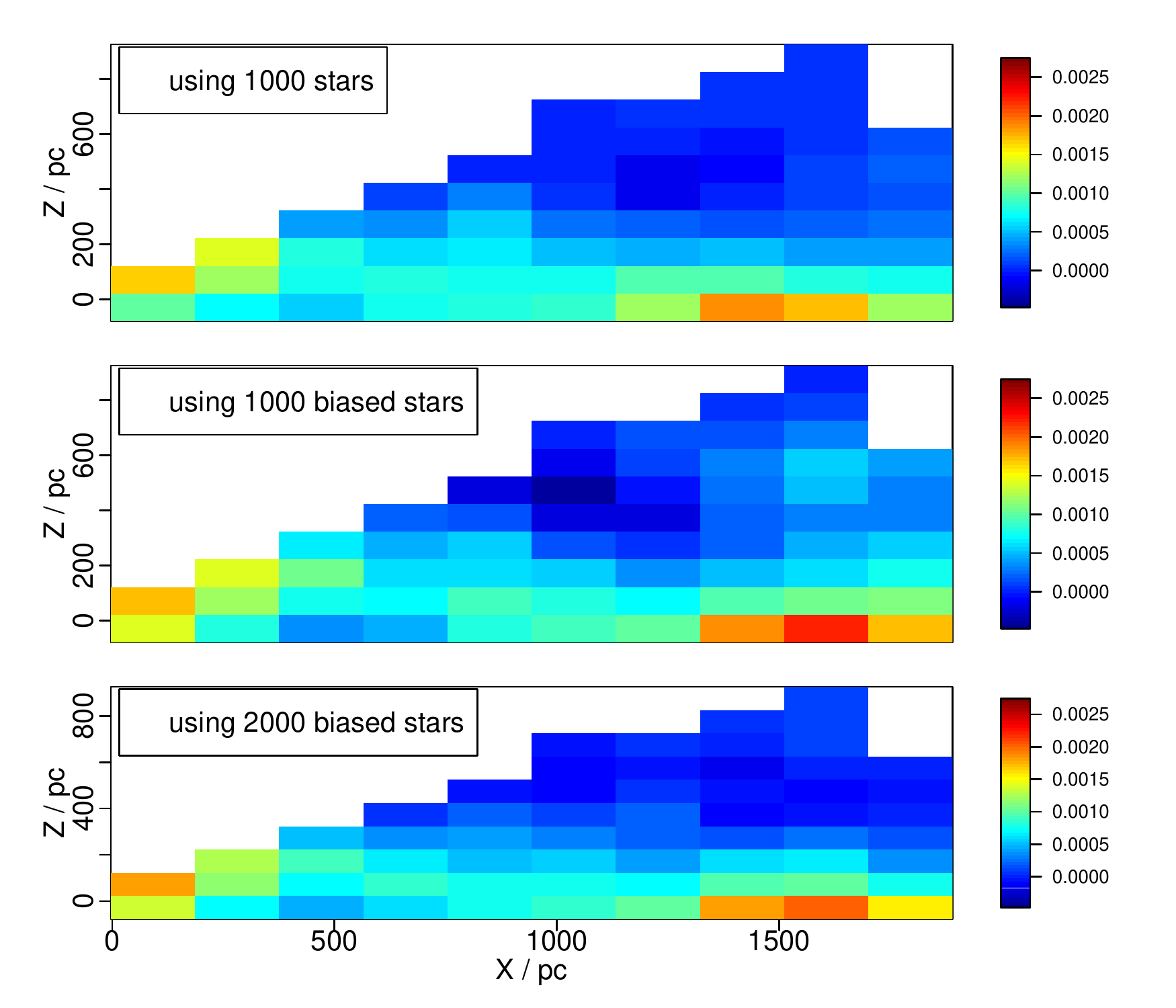}
\caption{Effect of different sampling on the dust density predictions (attenuation per parsec, shown as the colour scale) for the region of $5^{\circ} {\textless} l {\textless} 7^{\circ}$ and $-1^{\circ} {\textless} b {\textless} 30^{\circ}$ using GUMS catalogue (section \ref{GUMS}). Top panel shows the predictions using 1000 randomly sampled input data (from the entire 52\,000), the middle panel is the results using 1000 input stars which sampled in a biased way towards high value attenuations, and the lower panel shows predictions in case of the same biased sampling but using 2000 stars. The results are pretty much consistent, capturing similar trends.}
\label{fig:nlwb_sampling}
\end{center}
\end{figure}

In the above we selected just 1000 stars from a possible 52\,000 which met our selection criteria. 
How might this selection affect our inference?
The top panel of
Figure \ref{fig:nlwb_sampling} shows the same results as in Figure \ref{fig:pred_sample1_1000_posl} but now giving the average dust density in the rectangular region. The middle panel shows the results when we instead select 1000 stars
but biased towards selecting higher attenuations. For the biased selection, we sample from the data with the probability of being selected proportional to the rank of the sorted attenuation values.
The bottom panel is for 2000 stars selected in this same biased manner.
Overall we see a high level of consistency, although not surprisingly, these latter two data sets reveal larger densities in the higher extinction areas.
There is essentially no difference between using 1000 and 2000 stars, however.
This is good news for our method, because it has a poor computational time scaling with the number of data points.

\section{Application to APOKASC data}\label{real_data}

Having demonstrated the basic features of our model on simulated data, we now apply
it to a set of real data to construct a 3D dust map in a small region of the Galaxy.
\begin{figure} 
\begin{center}
\includegraphics[width=0.50\textwidth, angle=0]{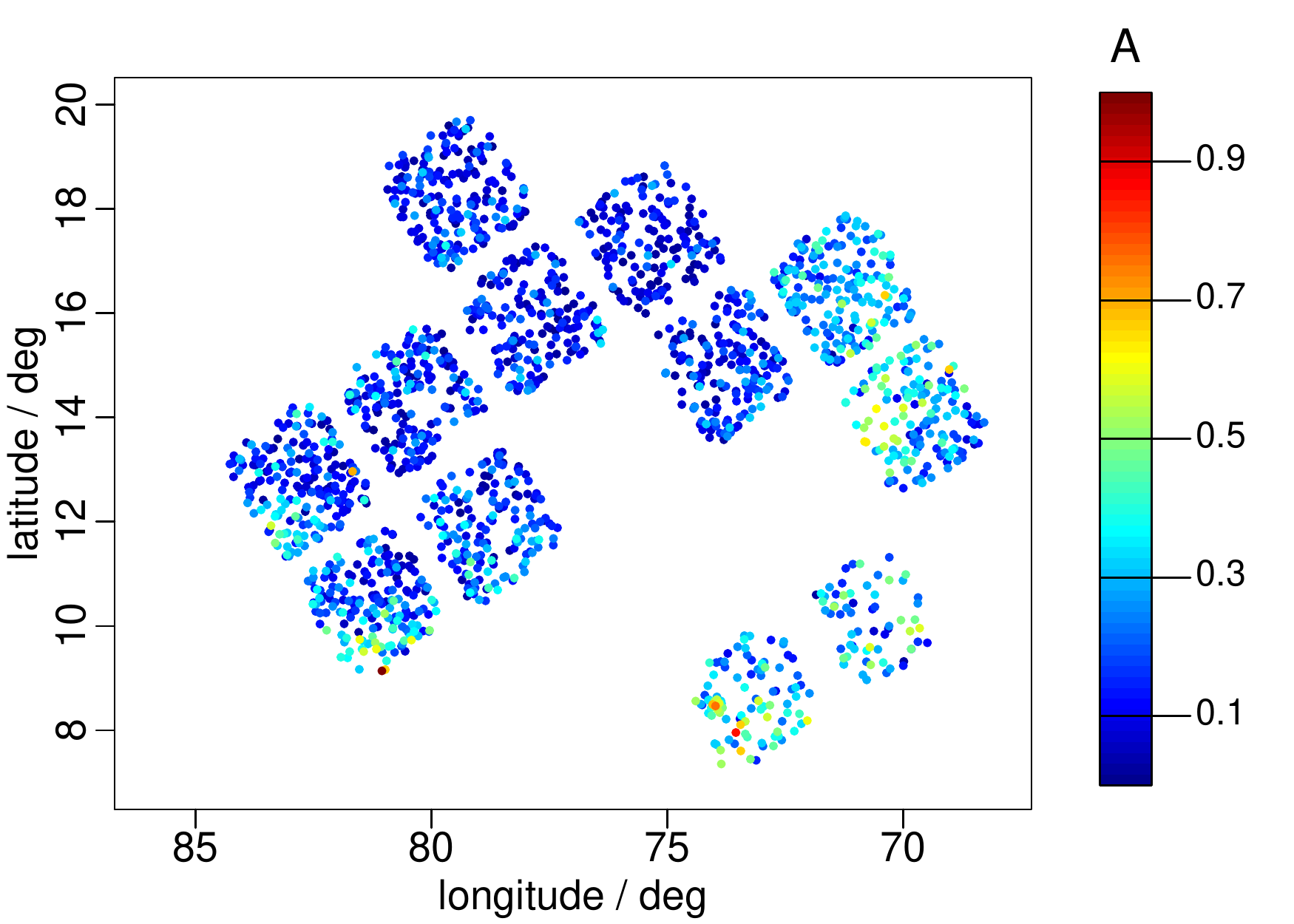}
\caption{
Extinction values of 1900 stars from \citet{2014MNRAS.445.2758R} lying within 3 kpc. The colour indicates the extinction in magnitudes. The discrete squares are due to the Kepler satellite fields.}
\label{fig:rod_data}
\end{center}
\end{figure}

We use extinctions and positions of nearly 2000 stars provided by \citet{2014MNRAS.445.2758R}. They use spectroscopic and asteroseismic
data of giants observed by APOGEE and the Kepler satellite (APOKASC catalogue), together with photometry from SDSS, 2MASS, and WISE and apply a Bayesian method to determine their extinctions and distances. The sample covers 17 degrees in longitude (from $68^{\circ}$ to $85^{\circ}$) and 14 degrees in latitude (from $6^{\circ}$ to $20^{\circ}$). We select stars with distance uncertainties of less than 10\%  within 3\,kpc and remove stars with negative extinctions. This leaves around 1900 stars with observed extinctions shown in figure \ref{fig:rod_data}. The spatial distribution arises from the Kepler's CCD placement in the focal plane. Many of the extinctions are small, but we do see patches of higher extinction on different scales, especially in the lower right of the plot.
\begin{figure}
\begin{center}
\includegraphics[width=0.50\textwidth, angle=0]{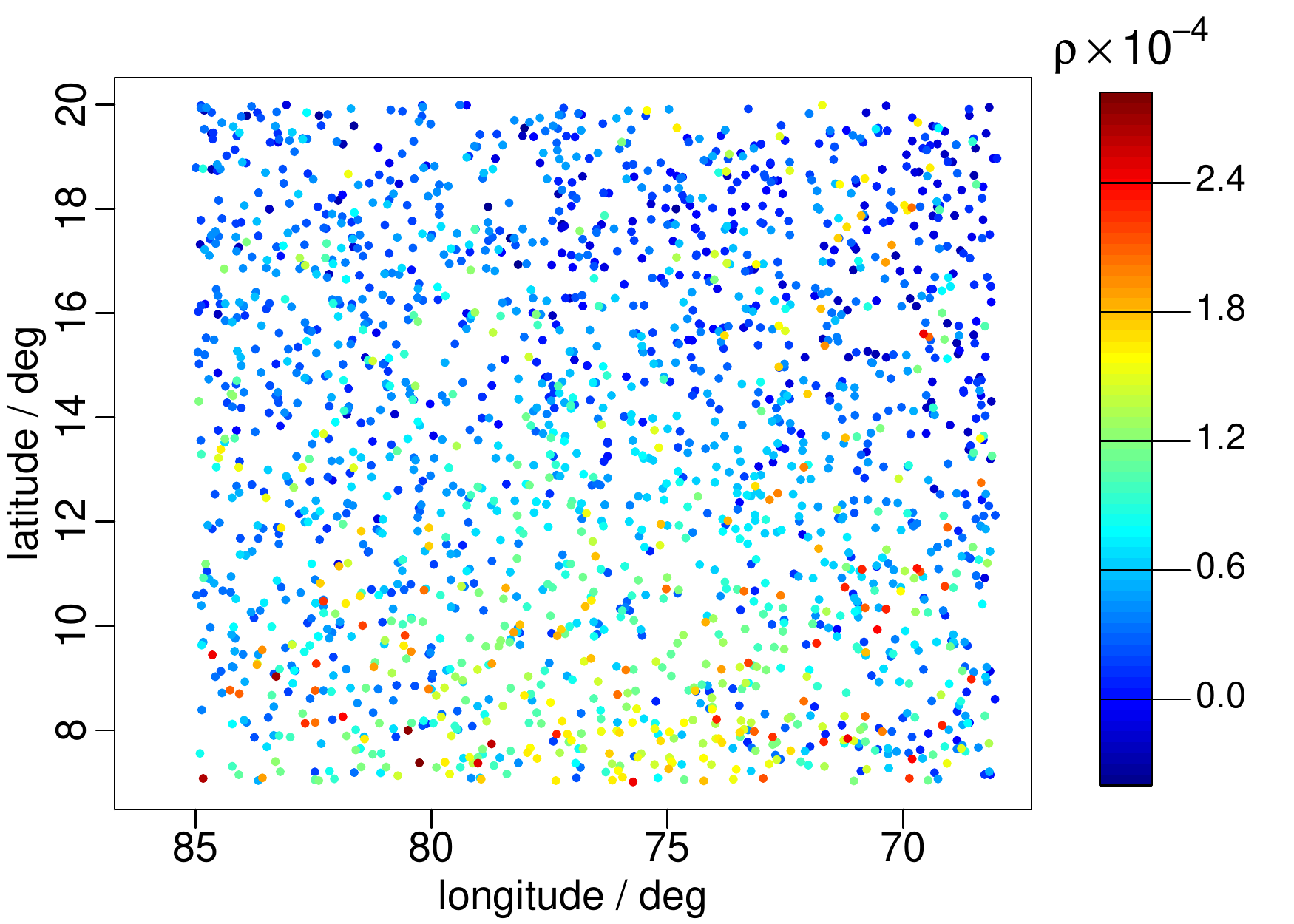}
\caption{Inferred dust density (attenuation per parsec) for the area of $68^{\circ} {\textless} l {\textless} 85^{\circ}$ and $6^{\circ} {\textless} b {\textless} 20^{\circ}$ out to 3 kpc using data from \citet{2014MNRAS.445.2758R} shown in 2D with the same frame as figure \ref{fig:rod_data} (the input data). It should be noted that dust density is local in 3D and these points lie at a range of distances as seen in figure \ref{fig:reconst_rod_res}.}
\label{fig:output_rod_2D}
\end{center}
\end{figure}

We apply our model to estimate the dust density (specifically: the mean and standard deviation of a Gaussian distribution at each point) at 2000 points distributed at random throughout the volume occupied by the stars. This has the shape of a pyramid with its apex at the Sun.  We use $\lambda$ = 2 kpc, $g$ = 250 pc, and $\theta$ = $3 {\times} {10}^{-7}$ (set as described in section \ref{simulated_data}).  Figure \ref{fig:output_rod_2D} shows the mean estimated dust density of each of these points projected onto the sky.  Comparing the coverage of this dust map with the input data makes it apparent that we can predict dust densities for points outside of the measured field. The model predicts an extended region of higher density at the bottom of the region, presumably driven by the higher extinctions for some stars in the lower fields in figure \ref{fig:rod_data}.

\begin{figure*}
\resizebox{\hsize}{!}{\includegraphics[clip=true]{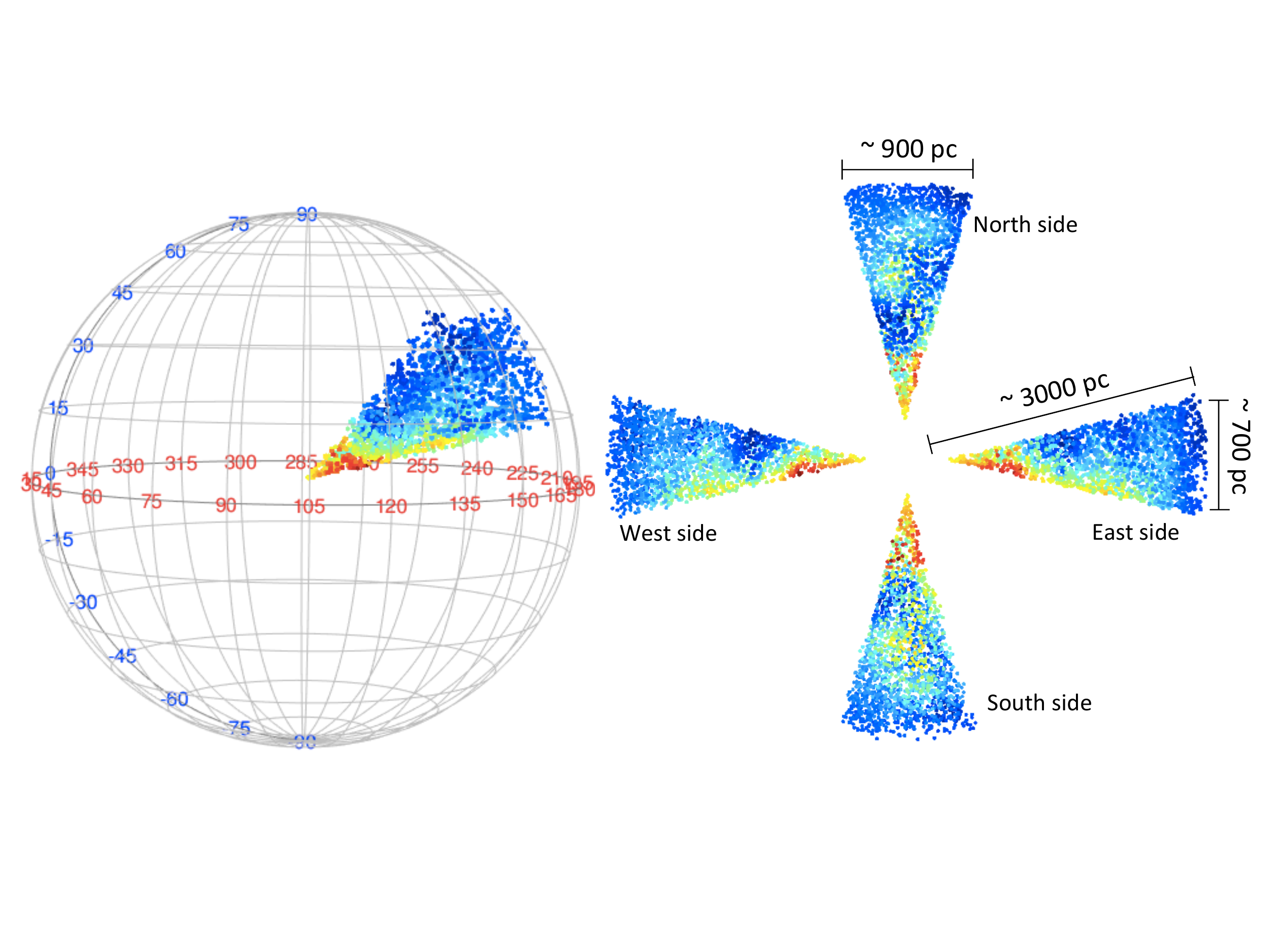}}
\caption{Dust densities estimated by our model over the pyramid-shaped region $68^{\circ} {\textless} l {\textless} 85^{\circ}$, $6^{\circ} {\textless} b {\textless} 20^{\circ}$, $d<3$\,kpc, computed using the extinction and distance data from \citet{2014MNRAS.445.2758R}. The left panel figure shows the data in three dimensions. The right panel shows the dust density in projection as viewed from the outside sides of the pyramid. 'north' means looking from top to bottom, perpendicular to the upper side of the pyramid, and similarly for 'south' (looking from the bottom). The 'east' and 'west' views are looking perpendicular to the sides of the pyramid, where 'east' means looking from higher longitudes to lower ones, and 'west' from lower to higher. Colour scale is as in figure \ref{fig:output_rod_2D}. It should be noted that these triangles are plotted with a larger scale parallel to their base (transverse to the l.o.s from the Sun) -- a larger opening angle than in reality -- in order to better resolve the details.  }
\label{fig:rod_pred}
\end{figure*}
As the inferred dust distribution covers a range of distances, a sky projection like Figure \ref{fig:output_rod_2D} 
does not give the full picture.
Figure \ref{fig:rod_pred} attempts to show the three-dimensional distribution. As expected from the input data, regions with higher dust densities are located at the lower latitudes, as is visible at the bottom of the east and west triangles.
There is also a relatively high density region at about 1.5\,kpc from the Sun which,
as it is most visible in the south triangle,
must be located predominantly in the southern part of the region. 
It is important to note that there is no l.o.s (``fingers of god'') effect in the dust reconstruction.
This is a direct consequence of the 
non-parametric nature of our model plus its minimal assumption on a smoothness prior.
This is one of the advantages that our approach brings over more traditional mapping techniques, which are generally based on independent estimates of the dust density along different l.o.s, which lead necessarily to the fingers of god.

One way to assess the performance of our model is to estimate the dust density at several points along the l.o.s towards to a star, and then to use equation \ref{eqn:dustsum} to predict the attenuation, $f_n$. The standard deviation in this prediction is found by taking the variance of equation \ref{eqn:dustsum},
\begin{equation}
{\rm Var}(f_n)\,=\, \gvec\trans \,\gpcov \, \gvec
\label{eqn:varA}
\end{equation}
where $\gvec$ is the the vector of cell sizes along that l.o.s, and $\gpcov$ is the covariance matrix (with elements given by equation \ref{eqn:covfunc}) of the dust densities in these cells. It should be noted that this expression takes into account the (often large) covariance between cells along a l.o.s.  (This is essential, because our model by its very nature assumes spatial correlations in the dust.)  We compute attenuation estimates and uncertainties in this way for 200 stars in the two lower right Kepler fields in figure \ref{fig:rod_data}.  For each star we use 15 cells with constant cells sizes per l.o.s. As the stars are at a side range of distances, this corresponds to cells sizes between 39 and 334\,pc.

\begin{figure}
\begin{center}
\includegraphics[width=0.50\textwidth, angle=0]{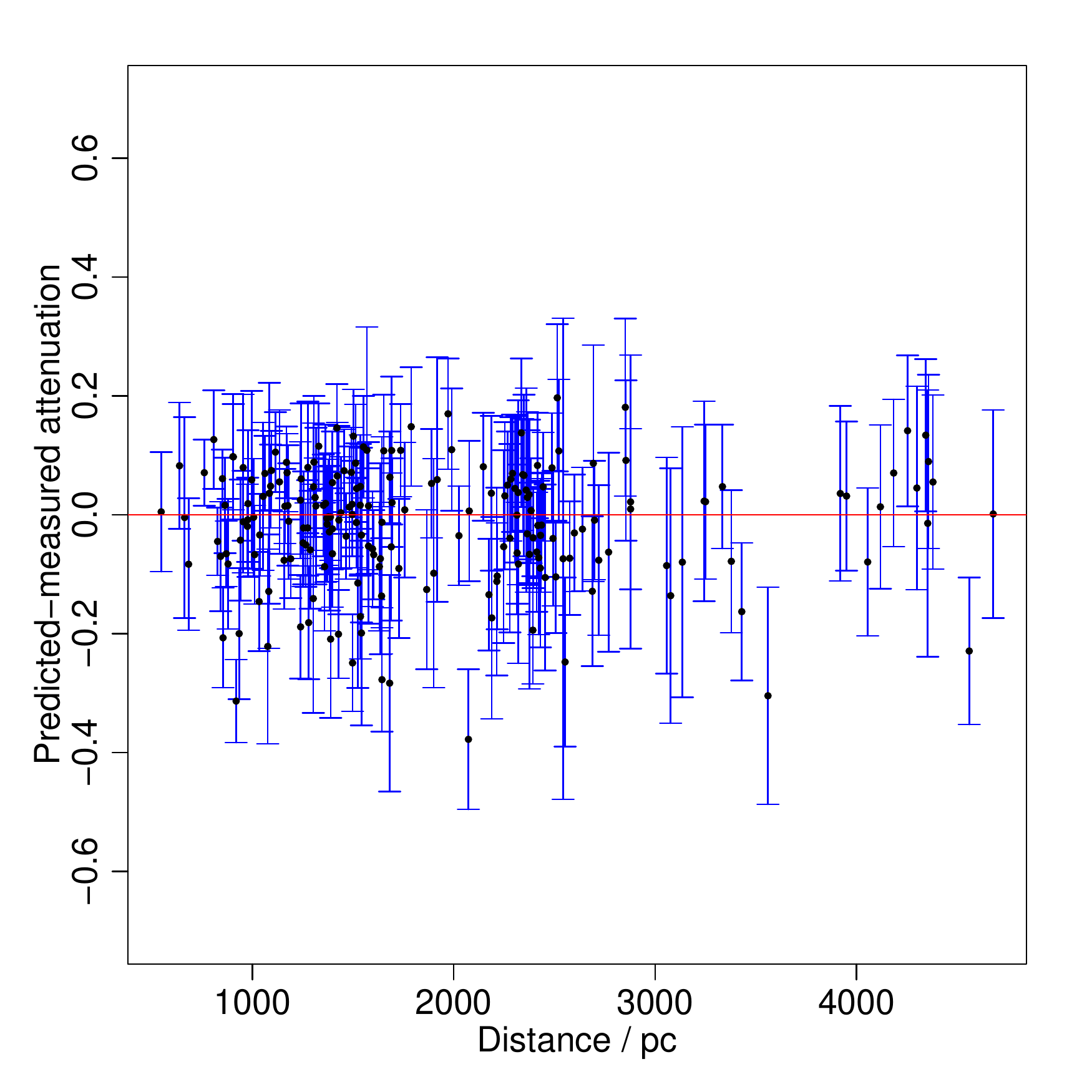}
\caption{Residuals between reconstructed and measured attenuations as a function of distance
for the 200 stars in the APOKASC data (from the bottom right two fields shown in Figure \ref{fig:rod_data}).
}
\label{fig:reconst_rod_res}
\end{center}
\end{figure}
\begin{figure}
\begin{center}
\includegraphics[width=0.50\textwidth, angle=0]{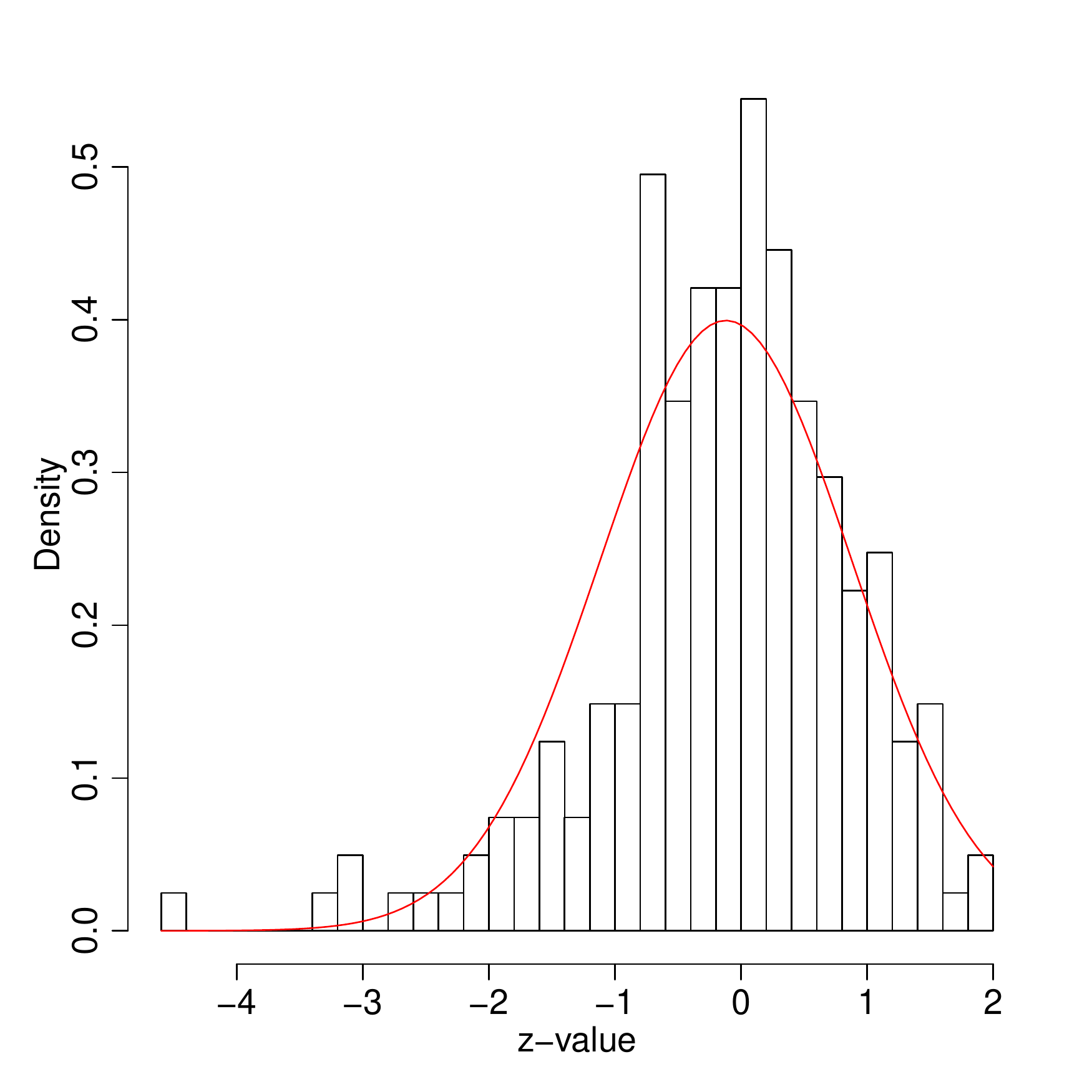}
\caption{Distribution of the scaled residuals, $(a_n - f_n)/\sigma$, where $\sigma^2 = \sigma_a^2 + {\rm Var}(f_n)$, for the set of 200 stars in the APOKASC data. The mean is -0.12 and the standard deviation is 0.99. The curve shows a Gaussian with these values for comparison.
}
\label{fig:reconst_rod_z}
\end{center}
\end{figure}

Figure \ref{fig:reconst_rod_res} shows the residuals (predicted minus measured) of the attenuations to these 200 stars, along with the corresponding error bars. These error bars are computed from the sum of the variance in the measurement, $\sigma_a^2$, and the variance from equation \ref{eqn:varA}.
We see no particular trend in the magnitude of either the residuals or their uncertainties with distance.
However, it looks as though there is a small negative bias, in the sense that our model slightly under estimates the measured attenuations. This is better seen in Figure \ref{fig:reconst_rod_z}, which shows the residuals scaled by their uncertainty estimates. A linear model with ideal Gaussian residuals would show a Gaussian distribution with zero mean and unit standard deviation. A negative bias is apparent. This is not necessarily a sign of a bad model, because the whole point of the model is to make an inference from the data subject to the smoothness constraint. Our model uses a zero mean prior for the dust density; thus, in the absence of data the model gives zero density. We can always add an offset to the covariance function of the Gaussian process to have a non-zero mean (e.g. if we want to examine the high density regions).

\begin{figure}
\begin{center}
\includegraphics[width=0.50\textwidth, angle=0]{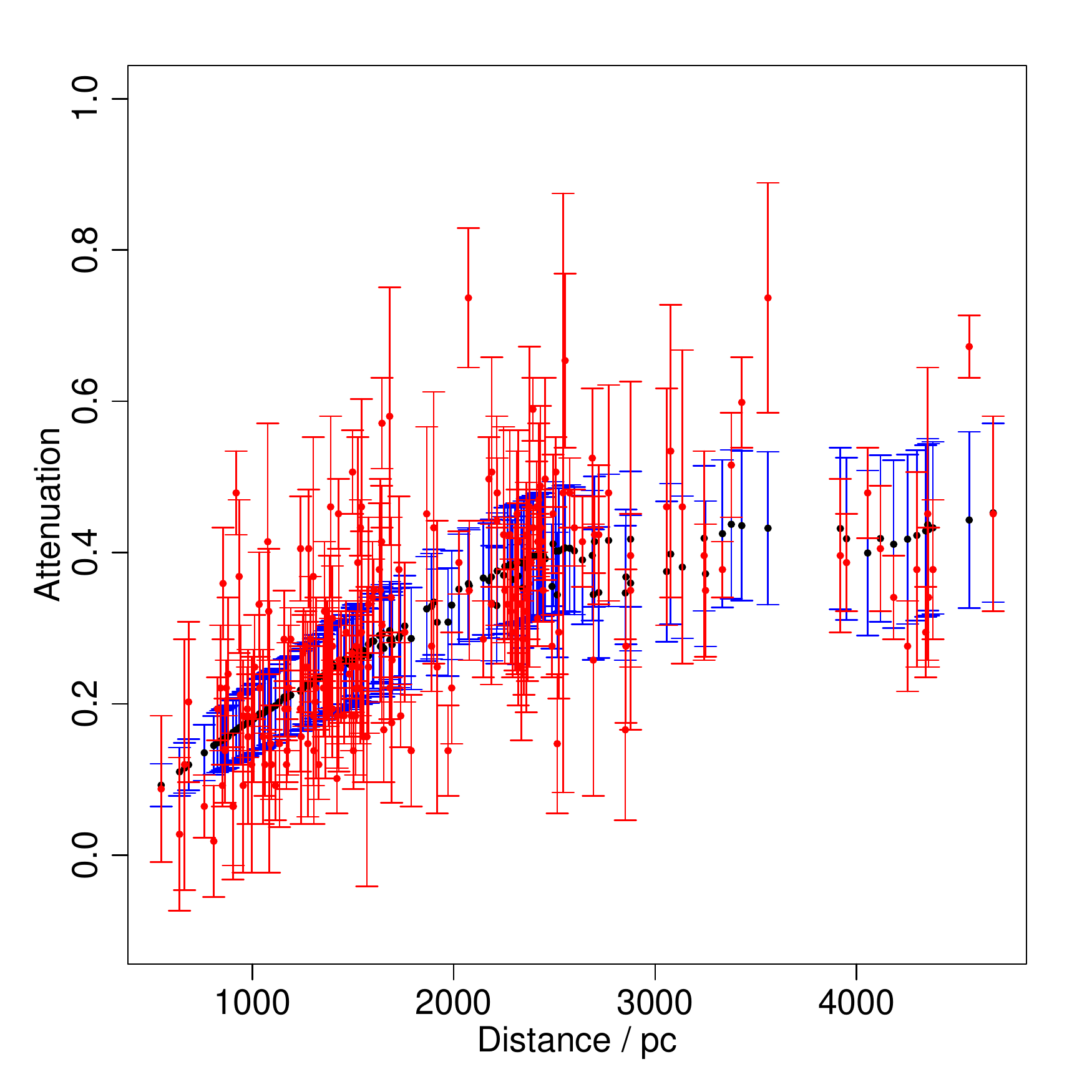}
\caption{Reconstructed attenuations as a function of distance (black dots with blue error bars) on top of the measured values (in red) for the set of 200 stars in the APOKASC data. It is important to note that the error bars in the predictions are highly correlated.
}
\label{fig:reconst_rod_dist}
\end{center}
\end{figure}

In contrast, the standard deviation is almost exactly unity, indicating that our Gaussian process is very good at estimating the uncertainty in its predictions.
Figure \ref{fig:reconst_rod_dist} shows the predicted values together with the measured values as a function of distance.
Recall that the error bars on the predictions are correlated. They increase with distance because the more distant stars have larger dust cells. There are a few measurements which lie well outside of the main envelope of the data, but given the number of points (and the size of their error bars in some cases) this is entirely consistent.
We see how the model has smoothed out these ``outliers'', on account of its built-in assumptions.
We further see that our model's underestimation of high extinctions is a consequence of its smooth variation with distance.

\section{Summary and future improvements}\label{discussion}

We have introduced a new non-parametric method for building a smooth, three-dimensional map of dust density which avoids l.o.s effects. It uses a Gaussian process prior to constrain the variation of the dust density in 3D space, but without assuming a specific functional form for the spatial dependence. It instead uses a covariance function which varies with the separation of points. This allows the model to infer the dust density in unobserved regions. Our model uses the 3D positions of stars together with their l.o.s extinctions as its input data and infers the posterior probability density function (PDF) over the dust at selected points. This PDF is a Gaussian, and we showed that its mean and standard deviation have analytic solutions.  While the l.o.s to the observed stars are divided into discrete cells, predictions are made at arbitrary points without any discretization being necessary.

We used a truncated covariance function in the Gaussian process which involves two hyperparameters: a correlation length scale $\lambda$ and a dust amplitude $\theta$. The latter can be set from the properties of the input data; adjusting $\lambda$ gives us flexibility to model dust variations on different length and amplitude scales.  The only requirement is that $\lambda$ be larger than the cell sizes.  While $\lambda$ is some characteristic length scale, our model can and does probe dust structures of much smaller scales (as in figure \ref{fig:mosaiclambdatheta}). 

We could try to fix these hyperparameters by calculating the Bayesian evidence. The evidence (or ``marginal likelihood'') is the probability of observing the data, for fixed $\lambda$ and $\theta$, averaged over all possible instantiations of the model. We compute this by drawing one sample from the J-dimensional Gaussian process prior (which gives us J values for the dust density), calculating the likelihood for these model dust densities, repeating it for a large number of times (e.g. $K = 10^5$), and then averaging these likelihoods 
\begin{equation}
P(\extvecN | \lambda, \theta) \,=\, \frac{1}{K} \sum_{k=1}^K {P}_{k}(\extvecN | \{{f}_{i}\}, {V}_{N})
\label{eqn:evidence}
\end{equation}
where $\{{f}_{i}\}$ are calculated attenuations (equation \ref{eqn:dustsumvec}) using dust densities drawn from the prior. Having done this for various $\lambda$ and $\theta$, we then calculate the Bayes factors, which are the ratio of these evidences (for different $\lambda$ and $\theta$) to the one with the specific values of $\lambda$ and $\theta$ used for our APOKASC data (section \ref{real_data}; $\lambda = 2$ kpc and $\theta$ = $4 {\times} {10}^{-7}$). We report these in table \ref{table:evidence}.

We get values for the Bayes factors in the case of APOKASC data (section \ref{real_data}) which agree broadly with what we calculated for $\lambda$ and $\theta$. But in the case of the simulated data (section \ref{simulated_data}), the Bayesian evidence does not give us a useful discrimination between models. Most values are very close to zero for a range of $\theta$ and $\lambda$ because our simulated data have high extinctions, which are not well represented by a Gaussian process prior with zero mean. The APOKASC data, in contrast, have smaller extinctions. This shows that using a non-zero mean in the Gaussian process prior will better construct the dust density in regions with higher extinctions, such as the disk of the galaxy and the spiral arms.
A different covariance function could also be used in the Gaussian process. We tested various forms of the covariance function, such as truncated exponential forms, but they did not make a significant difference to our results. The covariance function that we are using has the advantage that we can get different variation slopes by changing $\alpha$ (see fig. \ref{fig:gneiting}).
\begin{table}
\caption[]{Logarithm of the Bayes factors for APOKASC data for different ranges of $\lambda$ and $\theta$.}
\begin{tabular}{ ||c||c||c||c||c|| } 
 \hline
 \multicolumn{5}{|c|}{$\log_{10}$\,(Bayes factor)} \\
 \hline
 $\theta$ $\backslash$ $\lambda$ (pc)  & 500 pc & 1000 pc & 2000 pc & 3000 pc \\
 \hline
 $1{\times}10^{-8}$ &$ -1.41$ & $-0.87$ & $-0.43$ & $-0.21$ \\ 
 \hline
 $1{\times}10^{-7}$ & $0.05$ & $0.14$ & $0.15$ & $0.13$ \\ 
 \hline
 $1{\times}10^{-6}$ & $0.05$ & $-0.06$ & $-0.18$ & $-0.26$ \\ 
 \hline
 $1{\times}10^{-5}$ & $-0.40$ & $-0.48$ & $-0.65$ & $-0.69$ \\
 \hline
\end{tabular}
\label{table:evidence}
\end{table}

One drawback of our model is that its computation time increases non-linearly with the number of stars, $N$, and the number of cells, $J$. As explained in Appendix \ref{sec:acceleration}, the time-consuming part is the (one-off) inversion of the $J\times J$ covariance matrix $\gpcovJ$, which takes time $\ofo(J^n)$ to compute, where $n$ is typically $\lesssim 3$ but can be reduced to around 2.3 \citep{matinvert}, as well as various matrix inversions and multiplications taking time $\ofo(NJ^2)$, which must be done for every prediction.  For a problem with $N=230$ and $J=3203$, inverting $\gpcovJ$ took two minutes (using a single core on a modest AMD Opteron 6380 CPU).  Making predictions at multiple points can then be done in parallel: 200 predictions took 4 minutes with 40 cores, or 1.2 seconds per point. The computation time for more points is proportional to the number of points.
For a problem with $N=1000$ and $J=8185$,
it took 40 minutes to invert $\gpcovJ$, and 34 seconds per point to make new predictions for 1000 new points (again with 40 cores).
This is 28 times longer than the previous case, which agrees reasonably well with the $\ofo(NJ^2)$ scaling suggested above
(which gives $(1000\times 8185^2)/(200\times 3203^2) = 33$). 

The limiting factor when scaling this up to larger applications may be the memory rather than the run-time. For the case of $J=12\,000$, we needed 8 GB of RAM per core. This number is determined primarily by the number of cells, $J$, because the largest matrix has size $J \times J$. However, as we use sparse matrix methods and a truncated covariance function, the RAM required will not continue to grow as $J^2$. It is rather the density of cells in space, rather than the number of cells, which will ultimately drive the memory requirements.

Using the run-time numbers from above, and ignoring memory limitations, then with $N$=10\,000 and $J$=100\,000, the $\gpcovJ$ inversion takes around 30 days (with just one core; this could be accelerated if $\gpcovJ$ inversion is parallelized too). Then even with 10\,000 cores running for 30 days we could only make predictions at 30\,000 points. This (and $N$) is too small to build up a useful dust density map over a large volume of space.  One way to accelerate the computations is to use approximate matrix inversions, which can be done in time $\ofo(N^2)$.  Alternatively, instead of trying to model the entire volume in one, we could partition it into partially overlapping regions, solve for each separately, and then join them.  Thought is required to combine the overlapping regions without discontinuities, but recall that any two points separated by more than $\lambda$ are not connected by our model anyway, due to the truncated covariance matrix (which itself does not produce discontinuities). Optimal partitioning is an area for future investigation.

While our method takes into account the uncertainties in the extinction measurements, it does not yet make use of the distance uncertainties. This will be necessary in some practical applications, as even from Gaia more distant and/or fainter stars will have poor distance estimates from the parallaxes \citep[e.g.][]{2015PASP..127..994B}. Including distances in the likelihood model is one approach, but would make the solution non-analytic. We are currently exploring this approach and will report on it in a future paper.

Once these developments have been made and tested, the model will be ready to be applied to the Gaia data.

\section*{Acknowledgments}
We wish to thank David Hogg for useful discussions and suggestions. We'd also like to thank Rene Andrae, Hans-Walter Rix, and Stuart Sale, for their constructive comments.
This project is partially funded by the Sonderforschungsbereich SFB\,881 ``The Milky Way System'' of the German Research Foundation (DFG). 

\bibliographystyle{aa}
\bibliography{paper1}

\begin{appendix}

\section{Analytic solution of the integral}\label{sec:analytic_solution}

To solve the integral in equation \ref{eqn:rhointA}, we write it as
\begin{equation}
P(\rhoJp | \extvecN) \,=\, \frac{1}{Z} \int e^{-\psi/2} \, d\rhovecJ \ 
\label{eqn:rhointB}
\end{equation}
where $Z$ is a normalization constant, and where from equation~\ref{eqn:gpJ} (with $J \rightarrow J\mplus 1$) and equation \ref{eqn:likeN}
\begin{alignat}{2}
\psi \,&=\, \rhovecJp\trans \gpcovJp\inv \rhovecJp +  (\extvecN - G\rhovecJ)\trans \likecovN\inv (\extvecN - G\rhovecJ) \nonumber\\
\,&=\, \rhovecJp\trans \gpcovJp\inv \rhovecJp  +  \extvecN\trans\likecovN\inv\extvecN -  \rhovecJ\trans\geomat\trans\likecovN\inv\extvecN + \nonumber\\
&\,\,\qquad \rhovecJ\trans\geomat\trans\likecovN\inv\geomat\rhovecJ - \extvecN\trans\likecovN\inv\geomat\rhovecJ
\label{eqn:psiA}
\end{alignat}
with
\begin{equation}
\rhovecJp \,=\, \left[ \begin{array}{c} \rhovecJ \\ \rhoJp \end{array} \right] \ .
\end{equation}
As each term in equation \ref{eqn:psiA} is a scalar, and transposing a scalar leaves it unchanged, we see that the third and fifth terms are identical.

To evaluate the integral we need to separate $\rhovecJ$ from $\rhoJp$ in $\rhovecJp$. 
We first partition the inverse covariance matrix in the following way
\begin{equation}
\gpcovJp\inv \,=\, \begin{bmatrix} \mmatJ & \mvecJ \\ \mvecJ\trans & \mu \end{bmatrix} 
\label{eqn:partgpcovinv}
\end{equation}
where $\mmatJ$ is a $J\mby J$ matrix, $\mvecJ$ is a $J\mby 1$ vector, and $\mu$ is a scalar.
It is important to note that these are components of the inverted matrix, not components of $\gpcovJp$ which are then inverted.
We can then write
\begin{alignat}{2}
\rhovecJp\trans \gpcovJp\inv \rhovecJp 
 \,&=\, \left[ \begin{array}{cc} \rhovecJ\trans & \rhoJp \end{array} \right] 
\begin{bmatrix} \mmatJ & \mvecJ \\ \mvecJ\trans & \mu \end{bmatrix} 
\left[ \begin{array}{c} \rhovecJ \\ \rhoJp \end{array} \right] \nonumber\\
 \,&=\, \rhovecJ\trans\mmatJ\rhovecJ +  2\rhoJp\mvecJ\trans\rhovecJ + \mu\rhoJp^2 \ .
\end{alignat}
Substituting this into equation \ref{eqn:psiA} and gathering together terms gives
\begin{alignat}{2}
\psi \,&=\, \rhovecJ\trans(\mmatJ + \geomat\trans\likecovN\inv\geomat)\rhovecJ + 2(\rhoJp\mvecJ\trans - \extvecN\trans\likecovN\inv\geomat)\rhovecJ \nonumber\\
 \,&\, + (\extvecN\trans\likecovN\inv\extvecN + \mu\rhoJp^2) \ 
\label{eqn:psiB}
\end{alignat}
which is a quadratic expression in $\rhovecJ$. The last term is independent of $\rhovecJ$ so can be taken out of the integral, allowing us to write equation~\ref{eqn:rhointB} as
\begin{alignat}{2}
P(\rhoJp | \extvecN) \,&=\, \frac{1}{Z} \exp\left[-\frac{1}{2}\extvecN\trans\likecovN\inv\extvecN \right] \exp\left[-\frac{1}{2}\mu\rhoJp^2 \right] \nonumber\\
\,&\, \int \exp\left[-\frac{1}{2}\rhovecJ\trans\rmatJ \rhovecJ + \bvecJ\trans\rhovecJ \right] d\rhovecJ \ 
\label{eqn:rhointC}
\end{alignat}
where
\begin{alignat}{2}
\rmatJ \,&=\, \mmatJ + \geomat\trans\likecovN\inv\geomat  \label{eqn:rmatJdef} \\ 
\bvecJ \,&=\, \geomat\trans\likecovN\inv\extvecN - \rhoJp\mvecJ \ .
\label{eqn:bvecJ}
\end{alignat}
The integral is a standard one
\begin{alignat}{2}
\int \exp\left[-\frac{1}{2}\rhovecJ\trans \rmatJ \rhovecJ + \bvecJ\trans\rhovecJ\right] d\rhovecJ \nonumber\\
\,=\, \frac{(2\pi)^{J/2}}{|\rmatJ|^{1/2}}
\exp\left[\frac{1}{2}\bvecJ\trans\rmatJ\inv\bvecJ \right]
\label{eqn:stdint}
\end{alignat}
provided $|\rmatJ| > 0$.
Using this we can write equation \ref{eqn:rhointC} as
\begin{alignat}{2}
P(\rhoJp | \extvecN) \,&=\, \frac{1}{Z} e^{-\phi/2} \hspace*{1em} {\rm where} \nonumber\\
\phi \,&=\, \mu\rhoJp^2 - \bvecJ\trans\rmatJ\inv\bvecJ
\label{eqn:rhointD}
\end{alignat}
and we have absorbed all factors which do not depend on $\rhoJp$ into the normalization constant.
Substituting for $\bvecJ$ from equation \ref{eqn:bvecJ} this becomes
\begin{alignat}{2}
\phi \,&=\, - \extvecN\trans\likecovN\inv\geomat\rmatJ\inv\geomat\trans\likecovN\inv\extvecN + (\mu - \mvecJ\trans\rmatJ\inv\mvecJ)\rhoJp^2 + \nonumber\\
  &\,\,\qquad 2\extvecN\trans\likecovN\inv\geomat\rmatJ\inv\mvecJ\rhoJp \ .
\end{alignat}
The first term does not depend on $\rhoJp$ so can also be absorbed into the normalization constant. Putting this into equation \ref{eqn:rhointD} gives us
\begin{equation}
P(\rhoJp | \extvecN)  \,=\, \frac{1}{Z} \exp \left[ -\frac{1}{2}\alpha\rhoJp^2 - \beta\rhoJp \right] \ 
\label{eqn:rhointE}
\end{equation}
where
\begin{alignat}{2}
\alpha \,&=\, \mu - \mvecJ\trans\rmatJ\inv\mvecJ \nonumber \\
\beta \,&=\, \extvecN\trans\likecovN\inv\geomat\rmatJ\inv\mvecJ \ .
\label{eqn:auxtermsA}
\end{alignat}
By completing the square in the exponent
\begin{equation}
 -\frac{1}{2}\alpha\rhoJp^2 - \beta\rhoJp  \,=\, -\frac{\alpha}{2}\left(\rhoJp + \frac{\beta}{\alpha}\right)^2 + \frac{\beta^2}{2\alpha}
\end{equation}
and taking the term $\exp(\frac{\beta^2}{2\alpha})$ outside of the integral (and absorbing it too into
normalization constant), we see that $P(\rhoJp | \extvecN)$ is just a Gaussian with mean $-\beta/\alpha$ and variance $1/\alpha$, i.e.\
\begin{equation}
P(\rhoJp | \extvecN) \,=\, \sqrt{\frac{\alpha}{2\pi}} \exp \left[   -\frac{\alpha}{2}\left(\rhoJp + \frac{\beta}{\alpha}\right)^2 \right] \ .
\end{equation}

\section{Accelerating the matrix evaluations via matrix identities}\label{sec:acceleration}

We can simplify the expressions for $\alpha$ and $\beta$ and thereby reduce the number of matrix multiplications using some standard matrix identities. We first relate the components of the inverted covariance matrix (equation \ref{eqn:partgpcovinv}) to the components of the covariance matrix itself, which we partition as
\begin{alignat}{2}
\gpcovJp \,&=\, \begin{bmatrix} \gpcovJ & \kvecJ \\ \kvecJ\trans & \kJp \end{bmatrix} \ .
\label{eqn:partgpcov}
\end{alignat}
$\gpcovJ$ is the $J\mby J$ covariance matrix involving only the fixed data. $\kvecJ$ is the $J\mby 1$ vector with elements given by $\gpcovel_{j,J+1}$, i.e.\ the covariance between the fixed data and the new point. $\kJp$ is the scalar $\gpcovel_{J+1,J+1}$, the covariance of the new point with itself (i.e.\ its variance).
Recall that each element of the covariance matrix is determined by the covariance function, such as equation \ref{eqn:covfunc}.
A standard result of matrix algebra (e.g.\ Press et al.\ 1992) allows us to write the components of the inverted matrix as
\begin{alignat}{2}
\mmatJ \,&=\, \gpcovJ\inv + (\gpcovJ\inv\kvecJ)(\kJp - \kvecJ\trans\gpcovJ\inv\kvecJ)\inv(\kvecJ\trans\gpcovJ\inv) \\
\mvecJ \,&=\, -(\gpcovJ\inv\kvecJ)(\kJp - \kvecJ\trans\gpcovJ\inv\kvecJ)\inv \\
\mu \,&=\, (\kJp - \kvecJ\trans\gpcovJ\inv\kvecJ)\inv \ .
\end{alignat}
Temporarily defining
\begin{alignat}{2}
\hvecJ \,&=\, \gpcovJ\inv\kvecJ
\label{eqn:hvecJdef}
\end{alignat}
allows us to write the matrix and vector parts of the inverted covariance matrix as
\begin{alignat}{2}
\mmatJ \,&=\, \gpcovJ\inv + \mu\hvecJ\hvecJ\trans \\
\mvecJ \,&=\, -\mu\hvecJ
\end{alignat}
It should be noted that $\hvecJ\hvecJ\trans$ is an outer product, so produces a matrix of size $J\mby J$.
We now use the matrix inversion lemma (also known as the Woodbury matrix identity; e.g.\ Press et al.\ 1992) to write, from equation \ref{eqn:rmatJdef}, 
\begin{equation}
\rmatJ\inv \,=\, \mmatJ\inv - \mmatJ\inv\geomat\trans(\likecovN + \geomat\mmatJ\inv\geomat\trans)\inv\geomat\mmatJ\inv \ .
\label{eqn:rmatJinv}
\end{equation}
The Sherman--Morrison formula (a special case of the matrix inversion lemma) allows us to write
\begin{alignat}{2}
\mmatJ\inv \,&=\, \gpcovJ - (1 + \mu\hvecJ\trans\gpcovJ\hvecJ)\inv \mu\gpcovJ\hvecJ\hvecJ\trans\gpcovJ \nonumber\\
 \,&=\,  \gpcovJ - (1 + \mu\kvecJ\trans\gpcovJ\inv\kvecJ)\inv \mu\kvecJ\kvecJ\trans
\end{alignat}
where we have substituted for $\hvecJ$ from equation \ref{eqn:hvecJdef}.

All of the above allows us to simplify our expressions for $\alpha$ and $\beta$ in equation \ref{eqn:auxtermsA} by using
\begin{alignat}{2}
\rmatJ\inv \,&=\,\mmatJ\inv - \mmatJ\inv\geomat\trans(\likecovN + \geomat\mmatJ\inv\geomat\trans)\inv\geomat\mmatJ\inv \nonumber\\
\mmatJ\inv \,&=\, \gpcovJ - (1 + \mu\kvecJ\trans\gpcovJ\inv\kvecJ)\inv \mu\kvecJ\kvecJ\trans \nonumber\\
\mvecJ \,&=\, -\mu\gpcovJ\inv\kvecJ \nonumber\\
\mu \,&=\, (\kJp - \kvecJ\trans\gpcovJ\inv\kvecJ)\inv \nonumber\\
\kJp \,&=\, \gpcovel_{J+1,J+1} \nonumber\\
\kvecJ \,&=\, (\gpcovel_{1,J+1}, \gpcovel_{2,J+1}, \ldots, \gpcovel_{J,J+1})\trans \ .
\label{eqn:auxtermsB}
\end{alignat}
$\gpcovJ\inv$ takes time $\ofo(J^n)$ to compute, where $n$ is typically $\lesssim 3$, but must only be done once. As we use a truncated covariance function, $\gpcovJ$ is a sparse matrix, a fact which we exploit when inverting it. 
To compute the PDF at each new point, various other matrix inversions and multiplications must be performed, the longest of which takes time $\ofo(NJ^2)$, because $J \geq N$. 
It should be noted that $G$ is always sparse (when $J=N$ it is even diagonal, although this would give very poor resolution dust density maps). Because $\gpcovJ$ is sparse, $\mmatJ\inv$ is also sparse.

\end{appendix}
\end{document}